\documentclass{jpp}
\usepackage{epstopdf}
\usepackage{graphicx}
\usepackage{epsfig}
\usepackage{hyperref}
\usepackage{amsfonts, amssymb, amsmath}
\usepackage{color}
\newcommand{\uu}{{\bf u}} 
\newcommand{\vv}{{\bf v}} 
\newcommand{\kk}{{\bf k}} 
\newcommand{\rr}{{\bf r}} 
\newcommand{\RR}{{\bf R}} 
\newcommand{\zz}{{\bf \hat z}} 
\newcommand{\cc}{\color{black}} 

\shorttitle{SGS effects in magnetised plasma turbulence}
\shortauthor{B. Teaca et al.}
\title{Sub-grid-scale effects in magnetised plasma turbulence}
\author{
Bogdan Teaca\aff{1,2}  \corresp{\email{bogdan.teaca@coventry.ac.uk}}, Evgeny A. Gorbunov\aff{1}, Daniel Told\aff{3}, Alejandro {Ba\~n\'on Navarro}\aff{3}, Frank Jenko\aff{3}
}
\affiliation{
\aff{1} Coventry University, Coventry CV1 5FB, United Kingdom 
\aff{2} University of Craiova, 13 A.I. Cuza Street, 200585 Craiova, Romania
\aff{3} Max-Planck-Institut f\"ur Plasmaphysik, Boltzmannstr. 2, D-85748 Garching, Germany
}

\begin{document}
\maketitle
\begin{abstract}
In the present paper, we use a coarse-graining approach to investigate the nonlinear redistribution of free energy in both position and scale space for weakly collisional magnetised plasma turbulence. For this purpose, we use high-resolution numerical simulations of gyrokinetic (GK) turbulence that span the proton-electron range of scales, in a straight magnetic guide field geometry. Accounting for the averaged effect of the particles' fast gyro-motion on the slow plasma fluctuations, the GK approximation captures the dominant energy redistribution mechanisms in strongly magnetised plasma turbulence. Here, the GK system is coarse-grained with respect to a cut-off scale, separating in real space the contributions to the nonlinear interactions from the coarse-grid-scales and the sub-grid-scales (SGS). We concentrate on the analysis of nonlinear SGS effects. Not only that this allows us to investigate the flux of free energy across the scales, but also to now analyse its spatial density. We find that the net value of scale flux is an order of magnitude smaller than both the positive and negative flux density contributions. The dependence of the results on the filter type is also analysed. Moreover, we investigate the advection of energy in position space. This rather novel approach for GK turbulence can help in the development of SGS models that account for advective unstable structures for space and fusion plasmas, and with the analysis of the turbulent transport saturation. 
\end{abstract}

\section{Introduction}

Our understanding of turbulence in collisionless magnetised plasma has increased dramatically during the last decade. This was spearheaded by the need to predict transport coefficients in magnetic confinement fusion and to explain solar wind observations at scales smaller than the ion gyroradius ($\rho_i$). In both laboratory and astrophysical settings, the relevant micro-physics requires a kinetic theory description, and it involves dynamics in a position-velocity phase space. While a non-perturbative Vlasov-Maxwell approach is ultimately desired, various approximations make the problem more {\cc tractable} from a numerical perspective. In particular, the gyrokinetic (GK) approximation for strongly magnetised plasma requires only a five-dimensional phase space (see \S\ref{gksyst}), and is used mostly in magnetic confinement fusion studies~\citep{Krommes:2012p1373, Helander:2015p1977, Fasoli:2016p1866}. In the astrophysical context, while it neglects cyclotron resonance {and has limitations that need to be considered~\citep{Told:2016p1915}, GK theory} captures the crucial dynamics of three-dimensional kinetic Alfv\'en wave (KAW) turbulence {\citep{PhysRevLett.110.225002}}. For turbulence at scales larger than the gyroradius, {drift kinetic approximations can reduce the dynamics further and capture the problem in a four-dimensional space} \citep{Zocco:2011p2143, Hatch:2014p1639}. 

{In our current work, we look at KAW turbulence in the range of perpendicular scales ($\ell_\perp\sim 1/k_\perp$) found between the ion and the electron gyroradii, $\rho_i > \ell_\perp> \rho_e$ (see \S\ref{num} for details on parameters). The use of a GK representation is needed to account for the gyroaverage effects on the ions' dynamics ($k_\perp\rho_i>1$). 
The comparison between the ion and electron species also allows us to roughly see the qualitative difference between gyrokinetic and drift-kinetic approximations, as the gyroaverage effects on the electrons are negligible in this range of scales ($k_\perp\rho_e<1$). The work itself, which benefits from the different qualitative behaviours of the ion and electron species, explores in position space the configuration of the energy flux across scales and the spatial energy transport, as we will elaborate next.} 

In {classical} turbulence, energising a fluctuation leads to a redistribution of energy via nonlinear interactions. This redistribution can occur as a flux that cascades the energy across scales, or as a spatial advection of energy in position space. The analysis of the redistribution of energy in wave space ($\kk$) cannot track the spatial advection, while a real space analysis cannot account for fluxes across scales. To merge the two, a coarse-grained analysis can be performed, which consists in filtering the system in regard to a cut-off scale ($\ell_c\sim 1/k_c$) and then performing an analysis in real space. Doing so localises the nonlinear dynamics in both position and scale space {simultaneously,} and is particularly useful if inhomogeneities develop. Coarse graining the system allows us to separate the nonlinear dynamics into coarse-grid-scale and sub-grid-scale (SGS) effects. The large, coarse-grid-scales do not cause particular problems when accounting for turbulence numerically. The complications that appear in the study of turbulence are mostly due to the sub-grid-scales. {These complications are usually considered in the development of Large Eddy Simulations (LES) models. However, the scaling of SGS terms relates to the fundamental problem of smoothness of turbulence, including for kinetic plasma \citep{Eyink:2018p2109}. Being the first numerical study of its kind for kinetic plasma, this work will concentrate on the introduction of the definitions used and the presentation of qualitative numerical results.} 

In the current paper, using numerical solutions of GK turbulence (\S\ref{gksyst}), we study the effects of SGS on the energy flux across scales and across compact structures in the perpendicular direction to the magnetic guide field (coarse graining introduced in \S\ref{sec_cggk}). We make this distinction based on the explicit form of the coarse-graining filter. Definitions with appropriate spatial density in position space are used. This allows the analysis of the redistribution of free energy in position space in addition to scale space (see \S\ref{numanal}). While the analysis uses a straight magnetic guide field geometry and is done for KAW relevant turbulence, introducing these effects will be useful for tokamak modelling, {\cc even though we do not present such models here. Being able to track point-wise the flow of free energy, our approach can help} with the analysis of advective unstable structures \citep{Mcmillan:2018p2120}, plasma {\cc blob dynamics} \citep{Theiler:2009p1053}, and saturation mechanisms for turbulent transport \citep{Howard:2016p2161}. {\cc While in the current paper we do not perform a coarse graining in velocity space, accounting for the redistribution of free energy in position space can help} future works that deal with Landau damping in inhomogeneous turbulent mediums, or that probe the nature of kinetic plasma turbulence \citep{2019PhRvX...9c1037G}. {\cc Last, a real space analysis can help with the automatisation of nonlinear diagnostics via machine learning algorithms, by identifying first in position space and then tracking in phase space the most important structures or events of interest (e.g. reconnections) for a turbulent plasma.}

\section{The gyrokinetic system} \label{gksyst}

\subsection{Highlights of past works on gyrokinetic turbulence} \label{hig}
    
In classical fluid turbulence \citep{Frisch}, the energy cascade, the locality of interactions and the intermittency behaviour are considered standard problems of interest. While turbulence at kinetic scales inherits all of them, it also adds the phase space mixing problem (includes Landau damping) that affects which route in phase space is selected for the thermalisation of plasma fluctuations. In magnetised plasma, all {of} these problems can be {tackled} via the GK approximation \citep[for a review on the formal derivation of the general equations, see][]{Brizard:2007p11}.     
    
The GK approximation was instrumental in probing turbulence at sub-ion scales ($k\rho_i\!>\!1$). GK theory assumes low plasma frequencies compared to the ion cyclotron frequency and small fluctuation levels compared to background quantities to remove the particles' fast gyro-motion, effectively reducing the relevant phase space to five-dimensions. This approach was adopted by \citet{Howes:2006p1280} for the study of kinetic Alfv\'en waves (KAW) and their turbulent cascade in the dissipative range of the solar wind \citep{Howes:2008p1524, Howes:2008p1132, Howes:2011p1370}. Compared to the use of global background profiles in tokamak geometries \citep[see][]{Krommes:2012p1373}, the use of local background approximations in a straight-field magnetic geometry, typical for the study of KAW turbulence, simplifies the underlying dynamics. 

Following the recipe of classical turbulence, a generalised free energy that is conserved in the absence of collisions was identified for GK turbulence \citep[see][]{Howes:2006p1280, Schekochihin:2008p1034, Schekochihin:2009p1131}. With the idea of a free energy cascade in phase space, the concept of the nonlinear phase mixing for GK was introduced as well \citep{Schekochihin:2008p1034, Schekochihin:2009p1131}. The nonlinear phase mixing occurs in the direction perpendicular to the magnetic guide field, and it refers in particular to the {creation} of small-scale structures in velocity space due to the small-scale structure in position space. This {effect} results from the nonlinear interaction between the distribution function and the gyroaveraged potential fields. The gyro-average represents the effect of the fast gyro-motion on the slower dynamics captured by GK theory. In the electrostatic limit, the phase space cascade and the nonlinear phase mixing were studied extensively \citep{Tatsuno:2009p1096, Tatsuno:2010p1363, Plunk:2011p1357, Tatsuno:2012p1421} for ``two-dimensional'' GK turbulence \citep[i.e. neglecting parallel dynamics, see][]{Plunk:2010p1360}. For the five-dimensional GK system, while still in the electrostatic limit, the energy balance equation and the energy cascade problem was studied by \citet{BanonNavarro:2011p1350, BanonNavarro:2011p1274, Nakata:2012p1387} and later by \citet{Teaca:2014p1571, Cerri:2014p1757, Maeyama:2015p1736}. Measuring the intensity of the energetic exchanges with the increase in separation between scales, the locality of the nonlinear interactions {was} studied for electrostatic GK turbulence in \cite{Teaca:2012p1415, Teaca:2014p1571} and for the electromagnetic KAW case in \citet{Told:2015p1712, Teaca:2017p1989}. 
{While GK turbulence exhibits a strong nonlocal {\cc interaction} character, \citet{Teaca:2017p1989} found that the nonlocal contribution is superimposed on top of a classic asymptotically local contribution that depends only with the separation between scales, rather than substituting the classic local character altogether}. This is encouraging when considering modelling the SGS effects. Last, the intermittency problem was looked at in phase space for KAW turbulence {by} \citet{Teaca:2019p2154}, where the deviation from scale invariance was measured directly on the distribution functions.

In relation to the dissipation route for magnetised plasma fluctuations, \citet{Told:2015p1712} showed via a multi-species GK simulation of KAW turbulence {\cc at plasma $\beta =1$} that electrons dissipate most of the free energy at ion scales ($k_\perp\rho_i \sim 1$), while ions dissipate at small scales ($k_\perp \rho_i>1$). Later, \citet{Navarro:2016p1965} showed {\cc on the same data} that the electrons prefer parallel collisions, indicative of parallel linear phase mixing {\citep{Hammett:1992p1538, Kanekar:2015p1911}}, while ions enter into a fluid-like cascade in the perpendicular direction. The linear phase mixing problem is tied to the Landau damping problem for GK turbulence \citep{Tenbarge:2013p1730} and has a non-trivial {effect} on its structure character \citep{Teaca:2019p2154}. The balance between linear phase mixing in the parallel direction and the nonlinear cascade in the perpendicular direction was introduced for a drift-kinetic reduced model in \citet{Schekochihin:2016p2141}. The four-dimensional drift-kinetic models in question integrate over the perpendicular velocity, while retaining the information for $k\rho_i<1$ scale dynamics \citep[see also][]{Hatch:2014p1639}. The use of these models was helpful in showcasing the linear flux of energy across parallel velocity scales induced by linear phase mixing, and its suppression that leads to the fluidisation of the kinetic turbulent problem \citep{Meyrand:2019p2148}. {Last, depending on the plasma parameters (plasma-$\beta$ in particular), \citet{Kawazura:2019p2135} showed via hybrid GK simulations that ions can exhibit parallel or (fluid-like) perpendicular dissipation routes in phase space}.

While understanding turbulence is a goal in itself, in tokamak studies, turbulence is seen as a problem that overcomplicates the study of heat and particle transport by energising small-scale fluctuations compared to the scale of the dominant linear instabilities \citep{Gorler:2008p1393, Gorler:2008p1617}. To model the effect of these small scales on the nonlinear interactions at large scales, large eddy simulations (LES) have been adopted for GK turbulence \citep{Morel:2011p1339, Morel:2012p1390}, and were refined further in \citet{BanonNavarro:2014p1535}. To put it simply, LES models SGS effects. While extensively known in the field of turbulence \citep[see][and the references within]{Eyink:2006p13}, an SGS analysis for kinetic turbulence was introduced by \citet{Eyink:2018p2109}, where the entropy cascade was rigorously defined for a full Vlasov-Maxwell-Landau system and an upper bound {scaling} computed via functional analysis. Considering that velocity space integrals are performed in addition to position space ones, cancellation effects cannot be overlooked when computing the actual fluxes across scales. To what degree the upper bound estimates overshoot the real levels can only be determined numerically, and it is one of the questions we answer in the current paper for the GK system.

\subsection{{Plasma parameters and} numerical simulation details} \label{num}

{Depending on the geometry of the external magnetic guide field and the plasma regime, the GK equations can have an intricate or simple explicit form. Before introducing the GK equations, we start by presenting the main parameters for the plasma considered and list the numerical details used to solve the system in practice.} 

{In this study, we look at a proton-electron plasma that is weakly collisional and strongly magnetised, and which evolves in the presence of a straight magnetic guide field ($B_0\zz$). Proton (referred to as ion) and electron species are included with their real mass ratio of $m_{i}/m_{e}=1836$. The plasma $\beta_{i}\equiv8\pi n_{i}T_{i}/B_{0}^{2}=1$ is chosen to match solar wind conditions at 1 astronomical unit. The plasma background is assumed to exhibit an isotropic thermodynamic equilibrium with a temperature ratio of $T_{i}/T_{e}=1$.} The electron collisionality is chosen to be $\nu_{e}=0.06\, \omega_{A0}$ (with $\nu_{i}=\sqrt{m_{e}/m_{i}}\nu_{e}$), and $\omega_{A0}$ being the frequency of the slowest Alfv\'en wave in the system. {This allows for a KAW cascade.} 

{The system is solved numerically with the help of the Eulerian code {\sc GENE}~\citep{gene}.} The data used in this work is from the simulation presented in \citet{Told:2015p1712}, and it is briefly summarised in the following:  The evolution of the gyrocenter distribution is tracked on a grid with the resolution $\{N_{x}, N_{y}, N_{z}, N_{v_{\parallel}}, N_{\mu},N_{s}\}=\{768, 768, 96, 48, 15, 2\}$, where ($N_x,N_y$) are the perpendicular, $(N_z)$ parallel, $(N_{v_{\parallel}})$ parallel velocity, and $(N_{\mu})$ magnetic moment grid points, respectively. This covers a perpendicular dealiased wavenumber range of $0.2\le k_{\perp}\rho_{i}\leq 51.2$ (or $0.0047\leq k_{\perp}\rho_{e}\leq1.19)$  in a domain $L_x=L_y=10\pi\rho_i$ ($\rho_s= \sqrt{T_{s} m_{s}} c / e B$). In the parallel direction, a $L_z=2\pi L_{\parallel}$ domain is used, where $L_{\parallel} \gg \rho_i$ is assumed by the construction of GK theory. A velocity domain up to three thermal velocity units {($v_{T,s}=\sqrt{2T_s/m_s}$)} is taken in each direction. The fluctuations in the system are driven to a steady state via a magnetic antenna potential, which is prescribed solely at the largest scale and evolved in time according to a Langevin equation~\citep{tenBargecpc14}.

\subsection{The gyrokinetic equations}

For the system considered above, we use the $\delta f$-approach. The particle distribution function of each plasma species $s$ is split into a time constant background $F_{s}$ and a perturbed part $\delta f_{s}$, with $\delta f_{s}/F_s \ll1$. We consider a local approximation for $F_{s}$, which for constant background density $n_s$ and temperature $T_s$ (again, $v_{T,s}=\sqrt{2T_s/m_s}$) has the Maxwellian form,
\begin{align}
F_{s}({\vv})=\frac{n_s}{( v_{T,s}\sqrt{\pi})^3} \exp\bigg{(}-\frac{v_\|^2+v_\perp^2}{v_{T,s}^2}\bigg{)}\,.
\end{align}
In the presence of a strong external magnetic field compared to the fluctuating electromagnetic fields, the dynamics of the plasma become strongly anisotropic ($k_\|/k_\perp\ll1$). More importantly, particles develop fast cyclotron motions (of gyro-frequency $\Omega_s=\frac{q_sB_0}{m_s c}$) compared to the rest of the plasma dynamics ($\omega/\Omega_s\ll1$). Employing the guiding center coordinate ($\RR_s=x{\bf \hat x}+y{\bf \hat y}+z\zz$) transformation
\begin{align}
\RR_s=\rr + \vv(\theta)\times\zz/\Omega_s=\rr + \vv_\perp(\theta)\times\zz/\Omega_s\, , \label{Rdef}
\end{align}
for $\vv(\theta)=\vv_\perp(\theta)+v_\|\zz =v_\perp \sin(\theta){\bf \hat x}+v_\perp \cos(\theta){\bf \hat y}+v_\|\zz$, and integrating the dynamics over the gyrophase angle ($\theta$) allows us to reduce the dimension of the phase space by one, obtaining the five-dimensional gyrocenter phase space ($\RR_s, v_\|, v_\perp$) of GK theory. We can substitute the perpendicular velocity with the magnetic moment $\mu=m_s v^2_\perp/2B_0$. {While} we do this in practice, some relations are more transparent when utilising $v_\perp$.

For this simple case, considering $\delta f_{s}/F_s \sim B_\perp/B_0 \sim B_\|/B_0 \sim (cE_\perp/v_{T,s})/B_0 \sim k_\|/k_\perp \sim \omega/\Omega_s \sim \epsilon \ll 1$ as the the GK ordering, expanding all fields in powers of $\epsilon$ and keeping contributions up to the first order, the perturbed distribution function becomes\footnote{Formally this is obtained via a pull-back operation \citep{Brizard:2007p11} on the gyrocenter distribution function and has an intricate expression. Only for a Maxwellian background $F_s$ does $\delta f_s$ ends up having the simple form given by~(\ref{deltaf}). Assuming a Maxwellian form for the background distribution function provides a tremendous simplification of the GK system.} 
\begin{align}
\delta f_s(\rr, \vv,t) =  -  \frac{q_s \phi(\rr,t) }{T_s}F_s(v) + h_s(\RR_s,v_\|,v_\perp,t) \label{deltaf} \, ,
\end{align}
where we see a Boltzmann response contribution and a non-adiabatic part, $h_s(\RR_s,v_\|,v_\perp,t)$, which here is the effective {\it gyrokinetic distribution function}.

The systematic expansion of the Vlasov-Maxwell system gives rise to the GK equations {\cc (see \citet{Brizard:2007p11} for a general Hamiltonian derivation, or \citet{Howes:2006p1280, Schekochihin:2009p1131} for a simpler presentation appropriate in our case). For the first order contribution $h_{s}(x,y,z,v_{\parallel},\mu,t)$, the GK equations have the form}
\begin{align}
\frac{\partial h_s}{\partial t}+ \frac{c}{B_0}\big{\{} \langle \chi \rangle_{\RR_s},h_s\big{\}} + v_{\parallel}\frac{\partial h_s}{\partial z}=\frac{q_s F_{s}}{T_{s}}\frac{\partial\langle \chi \rangle_{\RR_s}}{\partial t}+\bigg{(}\frac{{\partial} h_s}{\partial t}\bigg{)\!}_c\ \label{GKeq} \ .
\end{align}

While the electromagnetic potentials are computed at the particle position ($\rr$), only their gyroaveraged contribution affect the GK dynamics. {\cc For clarity, the gyroaveraged gyrokinetic potential $ \langle \chi(\rr)   \rangle_{\RR_s} = \frac{1}{2\pi}\int_0^{2\pi}  \chi\big{(}\RR_s-{\vv_\perp(\theta)\times\zz}/{\Omega_s} \big{)} d\theta$ is found via its wave-space representation as}
\begin{align}
\langle \chi \rangle_{\RR_s}= \sum_{\kk} e^{i \kk \cdot \RR} \bigg{[} J_0(\lambda_s) \Big{(}\hat\phi({\bf k}) -\frac{v_\parallel}{c} \hat A_{\parallel}{(\bf k)} \Big{)}+\frac{2\mu}{q_s}\frac{J_1(\lambda_s)}{\lambda_s} \hat B_{\parallel}({\bf k})\bigg{]}\  ,
\end{align}
where $J_0(\lambda_s)$ and $J_1(\lambda_s)$ are zero and first order Bessel functions, with $\lambda_s=k_\perp v_\perp/\Omega_s=k_\perp\sqrt{ 2\mu B_0/m_s}/\Omega_s$. The first order self-consistent electrostatic potential ($\phi$), magnetic potential in the parallel direction ($A_\parallel$), and magnetic fluctuation in the parallel direction ($B_{\parallel}$) are obtained in wave space from their respective GK field equations as, 
\begin{align}
\hat \phi(\kk,t)&=\sum_s 2\pi q_s\frac{  B_0}{m_s}\int_{-\infty}^{+\infty}dv_\|\int_{0}^{+\infty}d\mu\, J_0(\lambda_s) \hat h_s(\kk, v_{\parallel},\mu,t) \bigg{/} \sum_s \frac{q_s^2 n_s}{T_s} \label{phik} \, ,\\
\hat A_\|(\kk,t)&=\frac{4\pi}{k_\perp^2 c} \sum_s 2\pi q_s\frac{  B_0}{m_s}\int_{-\infty}^{+\infty}dv_\|\int_{0}^{+\infty}d\mu\, v_\| J_0(\lambda_s)  \hat h_s(\kk, v_{\parallel},\mu,t) \label{Apark} \, ,\\
\hat B_\|(\kk,t)&=-{4\pi} \sum_s 2\pi \frac{  B_0}{m_s}\int_{-\infty}^{+\infty}dv_\|\int_{0}^{+\infty}d\mu\,    \mu \frac{2J_1(\lambda_s)}{\lambda_s} \hat h_s(\kk, v_{\parallel},\mu,t) \label{Bpark} \, .
\end{align} 
Considering the values of the Bessel functions $J_0$ and $2J_1(\lambda_s)/\lambda_s$, we see that the gyroaverage operation cannot be ignored for $k_\perp \rho_i >1$ scales, where we can imagine the problem as the distribution of a system of electrical charged rings. Conversely, the gyroaverage operation is not that important for $k_\perp \rho_i <1$ scales, and drift-kinetic approximations  can be obtained in the  $k_\perp \rho_i \ll1$ limit, {\cc which can still account for gyroaverage effects in a simplified way \citep{Hammett:1992p1538, Hatch:2014p1639}}.

The ${(}{\partial h_s}/{\partial t}{)\!}_c$ term represents the action of collisions, which are here modelled through the action of a linearised Landau-Boltzmann collision operator~\citep[see supplementary material from][]{Navarro:2016p1965}. Collisions represent the ultimate sink of plasma fluctuations and, in the collisionless limit, they are assumed to occur at very small scales in velocity space. For GK theory, due to the nonlinear phase mixing, the small scales in the perpendicular velocity and the perpendicular small scales in position space are linked. As a result, dissipation in the perpendicular direction occurs similarly as for a fluid via an effective (hyper) Laplacian term in position space. For GK turbulence, the break from the fluidisation can occur only when the parallel collisions dominate \citep{Navarro:2016p1965} {\cc and higher velocity moments in the $v_\|$ direction become excited via linear phase mixing.}

\begin{figure}
\centerline{
\includegraphics[width = 1\textwidth]{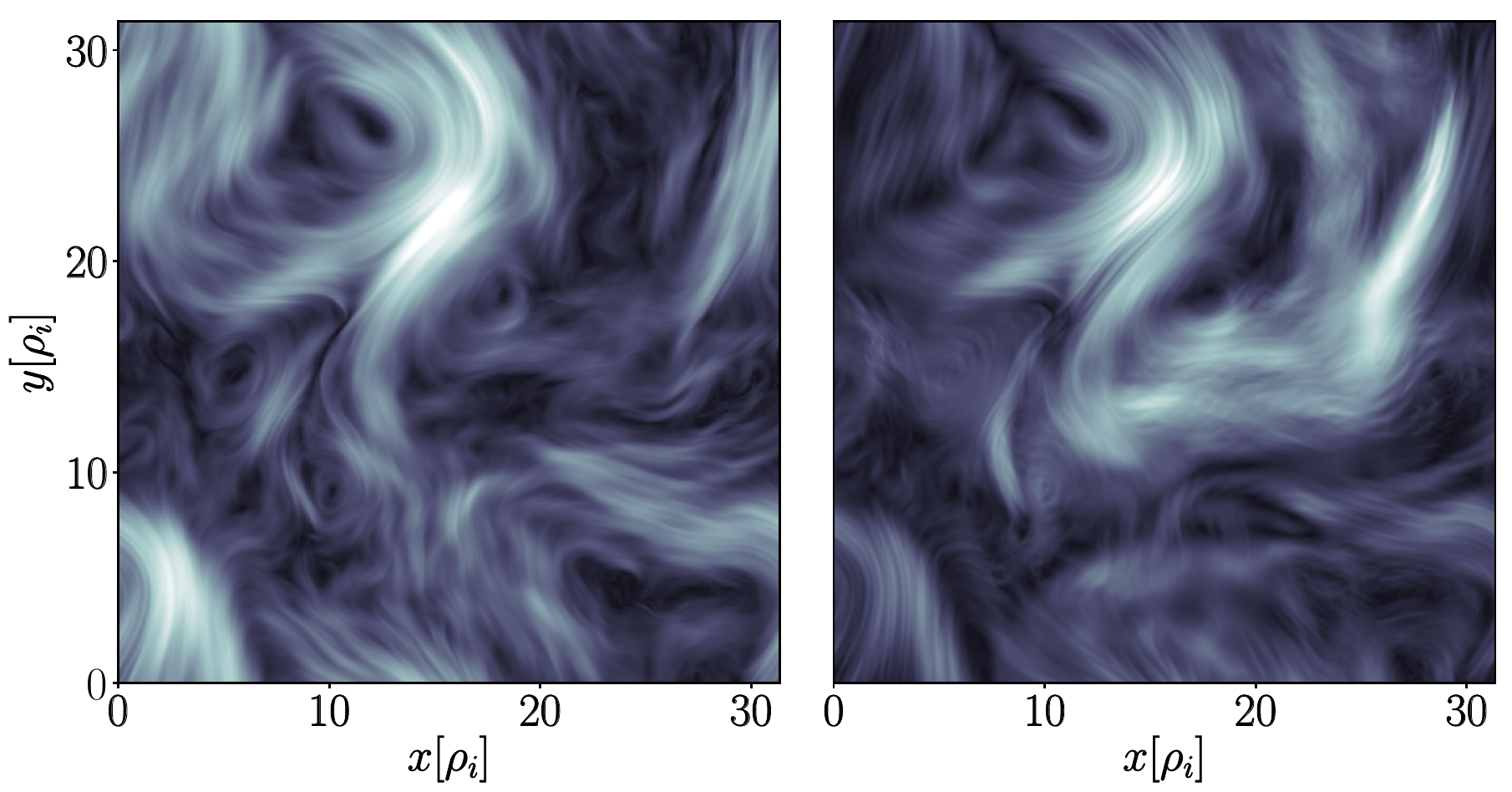}  }
\caption{Generalised drift velocity ${\bf u}_s$ for the GK system, at one point in $z$, $v_\|$ and $\mu$ {\cc ($v_\|=-0.31$ and $\mu=0.015$ in thermal velocity units)}. The difference between the ion and electron species consists in the gyroaverage of $\chi$. The electrons (left), which have a small gyroaverage that can be neglected for the scales depicted, show classical turbulent structures. For the ions (right), we see the phase mixing caused by the gyroaverage.}
\label{fig:driftv}
\end{figure}

The nonlinear structure is given in terms of the spatial Poisson bracket (to simplify the notation of gradients, from now on $\nabla\equiv\nabla_{\RR_s} ={\partial}/{\partial \RR_s}$),
\begin{align}
\big{\{}  a, b\big{\}}= [\nabla a \times \nabla b] \cdot {\zz}=\frac{\partial a}{\partial x}\frac{\partial b}{\partial y}-\frac{\partial a}{\partial y}\frac{\partial b}{\partial x}\, ,
\end{align}
which possesses all its properties (antisymmetry, bilinearity, etc. see Appendix \ref{appA} for details). To highlight 
the advective role of the nonlinearity, we can rewrite it as  
\begin{align}
\frac{c}{B_0}\big{\{} \langle \chi \rangle_{\RR_s},h_s\big{\}}= {\bf u}_s\cdot \nabla h_s=\nabla\cdot({\bf u}_s h_s) \, ,
\end{align}
where the advective velocity ${\bf u}_s$ is simply the generalised drift velocity for GK,
\begin{align}
{\bf u}_s=-\frac{c}{B_0}\Big{[}\nabla \langle \chi \rangle_{\RR_s} \times {\zz} \Big{]} \label{udef} \, .
\end{align}
By definition, it is clear that ${\bf u}_s={\bf u}_s(x,y,z,v_\|,\mu,t)$ is zero-divergent ($\nabla\cdot {\bf u}_s=0$) and that it differs slightly for each species due to the gyroaverage, see Figure~\ref{fig:driftv}. For the analysis of the nonlinear redistribution of free energy, we will utilise the advective velocity form for the nonlinear term. While key results will be presented in Poisson bracket form as well, the advective velocity form {\cc allows} for a much simpler connection with classical turbulence. {As mentioned in \S\ref{hig}, the {\cc analog} to the energy cascade in classical turbulence is given for GK turbulence by the free energy cascade.}

\subsection{The free energy}

\begin{figure}
\centerline{
\includegraphics[width = 0.8\textwidth]{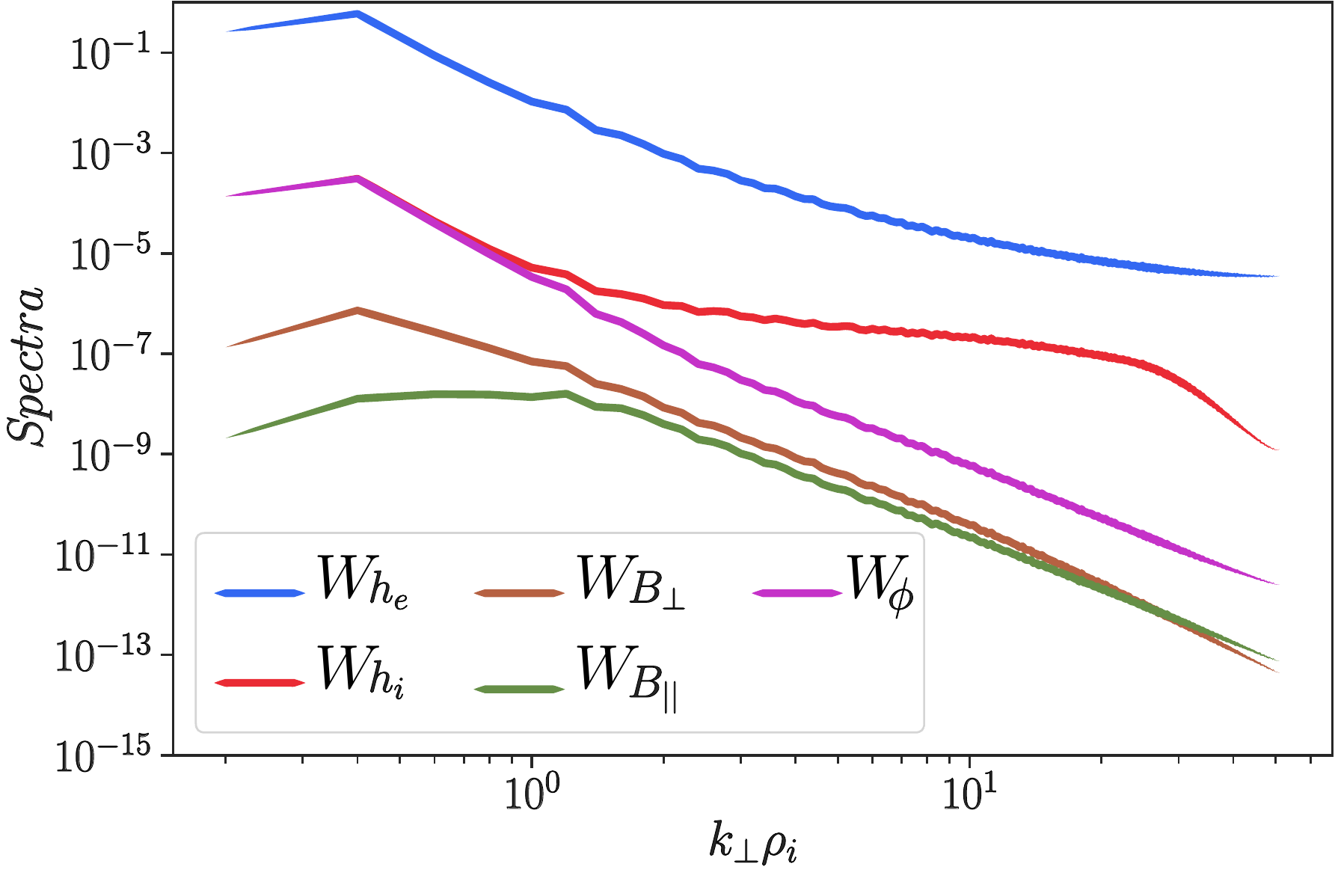}
}
\caption{Spectra in the perpendicular direction for the contributions to the free energy, normalised to the global level of free energy in the system.}
\label{fig:spectraW}
\end{figure}

As presented in \citet{Howes:2006p1280, Schekochihin:2008p1034, Schekochihin:2009p1131}, the generalised free energy is conserved for GK  turbulence in the absence of collisions and external sources. The free energy is defined as,
\begin{align}
W&=\int d^3r\,\bigg{[} \sum_s \int d^3v \frac{T_s  \delta f_s^2 }{2F_s} +\frac{B^2}{8\pi}\bigg{]},\label{feeq}
\end{align}
where we neglect the electric field energy contribution due to free charges, as the scales of interest here are much larger than the Debye length. Considering the quantities that express the GK equation and eq.~(\ref{deltaf}), the equivalent definitions are obtained,  
\begin{align}
W&=\int d^3r\,\bigg{[} \sum_s\int d^3v \frac{T_s  \langle h_s^2\rangle_\rr }{2F_s}  - \sum_s \frac{q^2_s n_s}{2T_s}\phi^2 +\frac{|\nabla_\perp A_\| |^2}{8\pi}+\frac{B_\|^2}{8\pi}\bigg{]} \nonumber \\
&=\sum_\kk \,\bigg{[} \sum_s \frac{2\pi B_0}{m_s}\int dv_\| d\mu \frac{T_s  }{2F_s}|\hat h_s(\kk,v_\|,\mu,t)|^2 \nonumber \\
&\qquad \qquad \qquad \qquad \qquad   - \sum_s \frac{q^2_s n_s}{2T_s}|\hat \phi(\kk)|^2 +\frac{k^2_\perp |\hat A_\|(\kk) |^2}{8\pi}+\frac{|\hat B_\|(\kk)|^2}{8\pi}\bigg{]}\, ,
\end{align}
{with $\langle h^2 \rangle_{\rr} = \frac{1}{2\pi}\int_0^{2\pi}  h^2\big{(}\rr+{\vv_\perp(\theta)\!\times\!\zz}/{\Omega_s}, v_\|, \vv_\perp(\theta) \big{)} d\theta  =\sum_{\kk} e^{i \kk \cdot \rr} J_0\big{(}\frac{k_\perp v_\perp}{\Omega_s}\big{)}\hat h^2(\kk, v_\|, v_\perp)$.

Considering the contribution of individual terms, with an appropriate selective summation in wave space defined as ${\sum_{\perp}\equiv\sum_{k_z} \sum_{|\kk_\perp| = k_\perp}^ {k_\perp + \Delta k} }$, we compute the unit band ($\Delta k$) spectra in the perpendicular direction
\begin{align}
W_{h_s}(k_\perp) &= \sum_{\perp} \frac{2\pi B_0}{m_s}\int dv_\| d\mu \frac{T_s  }{2F_s}|\hat h_s(\kk,v_\|,\mu,t)|^2 \, ,\\
W_\phi(k_\perp) &= \sum_{\perp} \sum_s \frac{q^2_s n_s}{2T_s}|\hat \phi(\kk)|^2\, , \\
W_{B_\perp}(k_\perp) &= \sum_{\perp} \frac{k^2_\perp |\hat A_\|(\kk) |^2}{8\pi} \, , \\
W_{B_\|}(k_\perp)&=  \sum_{\perp} \frac{|\hat B_\|(\kk)|^2}{8\pi} \, .
\end{align}
The total free energy spectrum can be found simply as the sum,
\begin{align}
W(k_\perp)=\sum_s W_{h_s}(k_\perp) - W_\phi(k_\perp) + W_{B_\perp}(k_\perp) + W_{B_\|}(k_\perp).
\end{align}
We plot in Figure~\ref{fig:spectraW} the spectra for all the contributions to the free energy. We see that the so-called \citep{Schekochihin:2008p1034} entropic contributions ($W_{h_s}$) dominate the free energy.} The scaling of the magnetic fields is the same as listed in \citet{Told:2016p1915}. Notably, for $k_\perp \rho_i<1$, $W_{h_i}$ and $W_{\phi}$ have the same energy, as expected in the MHD limit \citep{Howes:2006p1280}.

From the GK equations (\ref{GKeq}), multiplying by $T_s h_s/F_s$ we obtain the balance equation for the $h_s^2$ variance,
\begin{align}
\frac{T_s}{2F_s}\bigg{[}\frac{\partial h^2_s}{\partial t}+ \frac{c}{B_0}\big{\{} \langle \chi \rangle_{\RR_s},h^2_s\big{\}} + v_{\parallel}\frac{\partial h^2_s}{\partial z}\bigg{]}={q_s h_s}\frac{\partial\langle \chi \rangle_{\RR_s}}{\partial t}+\frac{T_s}{F_s}h_s\bigg{(}\frac{{\partial}h_s}{\partial t}\bigg{)\!}_c\ \label{GKeq2} \ .
\end{align}
Integrating over the velocity space, position and summing over all species we can show that we recover the evolution of the free energy (see Appendix~\ref{app_energy}),
\begin{align}
\frac{dW}{dt}=\frac{d}{dt}\int d^3r\,\bigg{[} \sum_s\bigg{(}\int d^3v \frac{T_s \langle h_s^2 \rangle_{\rr}}{2F_s} - \frac{q^2_s\phi^2n_s}{2T_s}\bigg{)}+\frac{B^2}{8\pi}\bigg{]}.
\end{align}
We are interested in the nonlinear contribution to the evolution of free energy for a scale, knowing that {\cc for a finite-scale system}, globally, nonlinear interactions conserve $h_s^2$ (see Appendix \ref{appA}),
\begin{align}
\frac{d W}{dt}\bigg{|}_{NL}=\int d^3R_s \sum_s\int d^3v \frac{T_s}{2F_s}\frac{c}{B_0}\big{\{} \langle \chi \rangle_{\RR_s},h^2_s\big{\}}=\int d^3R_s \sum_s\int d^3v \frac{T_s}{2F_s} \uu_s\cdot \nabla h_s^2=0\, .
\end{align}
Next, we look at the GK equations and at the nonlinear contribution to the free energy balance for a system coarse-grained in the perpendicular direction in gyrocenter space.

\section{The coarse-grained gyrokinetic system} \label{sec_cggk}

\subsection{Definition of coarse graining}

{The coarse graining of the Vlasov-Maxwell kinetic system was done by \citet{Eyink:2018p2109} using isotropic kernels assumed to be smooth (e.g. infinitely differentiable) and rapidly decaying (e.g. compact) phase space functions. For magnetised plasma turbulence captured by GK theory, the parallel and perpendicular scales are too disjointed in size to justify the use of isotopic filtering kernels. We concentrate here on the perpendicular scales. Accounting that the GK dynamics of interest occur in the gyrocenter space, we define the perpendicular coarse-grid filtering for a $\RR_{\perp s}$ function as
\begin{align}
\overline{a}(\overline \RR_{\perp s}) &= \int d \RR'_{\perp s} G_\ell(\RR'_{\perp s}) a(\overline \RR_{\perp s}+\RR'_{\perp s}) \\
&= \int d \RR'_{\perp s} G_\ell( \RR_{\perp s}'-\overline \RR_{\perp s}) a(\RR'_{\perp s})=[G_\ell\star a](\overline \RR_{\perp s}) \, .
\end{align} 
The $\star$ symbol denotes the convolution operation. The filtering functions are considered as $G_\ell(\RR_{\perp s})=\ell_c^{-3}G(\RR_{\perp s}/\ell_c)$ with the $G$ kernels having a series of desirable properties,
\begin{align}
G(\RR_{\perp s})&\ge0 \ \ \mbox{(non-negative)}, \label{gp1}\\
\int d \RR'_{\perp s}\, G(\RR_{\perp s}) &=1 \ \ \mbox{(normalised)}, \\
\int d \RR'_{\perp s}\,\RR_{\perp s} \,G(\RR_{\perp s}) &={\bf 0}\ \ \mbox{(centered)}, \\
\int d \RR'_{\perp s}\, |\RR_{\perp s}|^2G(\RR_{\perp s}) &=1\ \ \mbox{(unit variance)} \label{gp4}.
\end{align} 
}

\begin{figure}
\centerline{\includegraphics[width = 0.75 \textwidth]{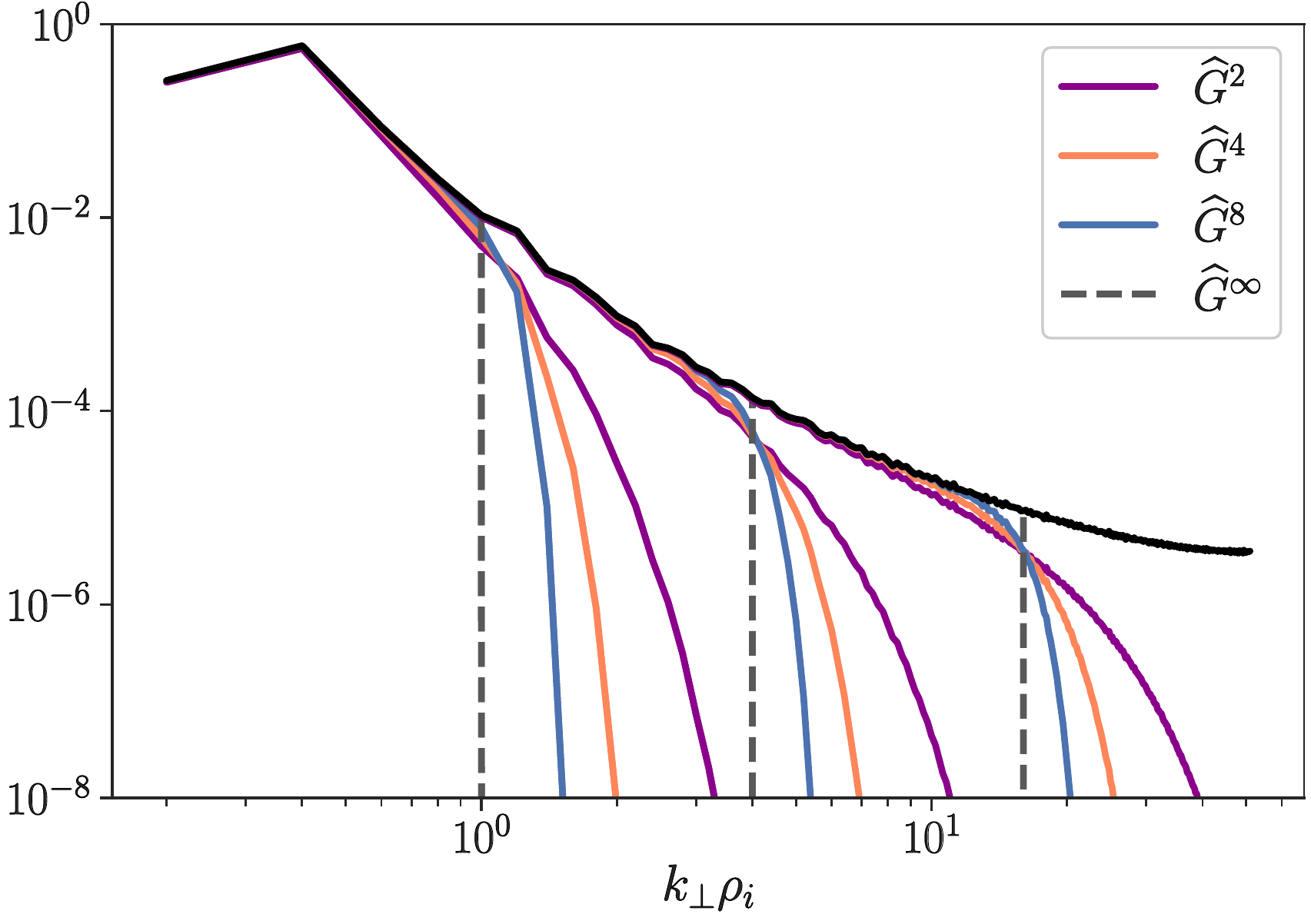}  }
\caption{Showcasing {\cc for $W_{h_e}$}, the wavenumber support for different filtering kernels $\hat G^\alpha$ considered for the same three cut-offs $k_c \rho_i=\{1,4,16\}$, with $\alpha=\{2,4,8\}$ and $\alpha=\infty$ representing the sharp $k$-filter. Not shown, for $\alpha=64$ the hyper-Gaussian filter would overlap near-identically the sharp filter.}
\label{fig:spec-filter}
\end{figure}

A Gaussian kernel,
\begin{align}
G(\RR_{\perp s}/\ell_c)=\frac{1}{\pi\ell_c^2}\exp\bigg{(}{-\frac{\RR_{\perp s}^2}{\ell_c^2}}\bigg{)}\, ,
\end{align} 
represents a good selection for the filtering function, as it obeys the properties~(\ref{gp1}-\ref{gp4}). This has the advantage of a simple wave space representation, $ \hat G(\kk_\perp/k_c)\sim e^{-(\kk_\perp/k_c)^2}$, which reduces the filtering convolution for the wavenumber cut-off {$k_c=2\pi/\ell_c$} to a simple multiplicative operation. However, in our work we will also consider a sharp $k$-filter in wave space {\cc (i.e. a Dirichlet kernel in real space)}. We also consider a general ({\it hyper-Gaussian}) kernel, 
\begin{align}
\hat G^\alpha(k_\perp/k_c) \sim e^{-(k_\perp/k_c)^\alpha}\,, \label{Galpha}
\end{align} 
that is isotropic in the perpendicular direction, knowing that for $\alpha=2$ we recover the Gaussian kernel and for large $\alpha$ we tend towards the sharp $k$-filter in respect to $k_\perp$. Figure \ref{fig:spec-filter} showcases this for the $W_{h_e}$ spectra, i.e. we filter the $h_e$ before computing the spectra. 

\subsection{Coarse-grained GK equations}

We start from the GK equations given in the advective velocity form and apply the coarse-graining operation ($G_\ell\star$) term by term. The overbar notation is moved only on the quantities that undergo coarse graining to obtain,
\begin{align}
\frac{\partial \overline{h}_s}{\partial t}+ \nabla \cdot ( \overline{{\bf u}_s h_s}) + v_{\parallel}\frac{\partial \overline{h}_s}{\partial z}=\frac{q_s F_{s}}{T_{s}}\frac{\partial\overline{\langle \chi \rangle}_{\RR_s}}{\partial t}+\bigg{(}\frac{{\partial}\overline{h}_s}{\partial t}\bigg{)\!}_c \ .
\end{align}
The field equations are linear in $h_s$ and thus do not pose any complications under the coarse-graining operation. Simply replacing $h_s$ by $\bar h_s$ in (\ref{phik}-\ref{Bpark}) {yields} the coarse-grained field equations.  

Natural for a nonlinear system, the $\overline{{\bf u}_s h_s}$ term, coarse-grained on a grid of resolution $\ell_c$, contains contributions from sub-grid-scales. In the nonlinear term, to separate the purely coarse-grained contributions from any SGS contributions, we make use of the cumulant 
\begin{align}
\overline{\tau}_s=\overline{{\bf u}_s {h_s}} -\overline{\bf u}_s \overline{h}_s\ .
\end{align}
Now $\overline{\tau}_s$ is the term that contains all the SGS contributions to the nonlinear dynamics. {An important property of $\overline{\tau}_s$ is that it is Galilean invariant by definition. Indeed, for $\uu'_s = \uu_s + {\bf U}$, with $\overline {\bf U}={\bf U}$, we have $\overline{\tau}'_s=\overline{({\bf u}_s +{\bf U})  {h_s}} -(\overline{\bf u}_s + \overline {\bf U}) \overline{h}_s\ =\overline{{\bf u}_s {h_s}} -\overline{\bf u}_s \overline{h}_s + {\bf U}  \overline{h}_s-{\bf U}  \overline{h}_s=\overline{\tau}_s$. In a similar way, $\overline{\tau}_s$ is also invariant to a $h'_s = h_s +H$ transformation, for $\overline {H}={H}$.}

The nonlinear term simply becomes,
\begin{align}
\nabla \cdot ( \overline{{\bf u}_s h_s})=\nabla \cdot ( \overline{\bf u}_s \overline{h}_s + \overline{\tau}_s) \label{nldef}\,,
\end{align}
where we now separate the coarse-grid-scale $\overline{\uu}_s$ and $\overline{h}_s$ from the sub-grid-scale $\overline \tau_s$ contributions.

Last, considering the definition (\ref{udef}) for $\uu_s$, we obtain the equivalent $\overline{\tau}_s$ formula, 
\begin{align}
\overline{\tau}_s=-\frac{c}{B_0}{\zz} \times \Big{[}\overline{ \nabla \langle \chi \rangle_{\RR_s} h_s} - \overline{\nabla \langle \chi \rangle}_{\RR_s} \overline{h}_s \Big{]},
\end{align}
which gives the SGS contribution to the nonlinear term expressed in term of the Poisson bracket structure as,
\begin{align}
\nabla \cdot \overline{\tau}_s=\frac{c}{B_0} \Big{[} \{\overline{ \langle \chi \rangle_{\RR_s} ,h_s }\} - \{ \overline{\langle \chi \rangle}_{\RR_s}, \overline{h}_s\} \Big{]}.
\end{align}

\subsection{The coarse-grained redistribution of free energy}
 
We look at the evolution of the coarse-grained free energy as result of the nonlinear interactions. This is obtained from (\ref{nldef}) by multiplying with $T_s\overline h_s/F_s$, integrating over the velocity and position space and summing over the species,
\begin{align}
\frac{d\overline W}{dt}\bigg{|}_{NL}= \sum_s \int d^3R_s\int d^3v \frac{T_s}{F_s}\bigg{[}\nabla \cdot ( \overline{\bf u}_s \overline{h}_s + \overline{\tau}_s)\bigg{]}\overline{h}_s = \sum_s \overline \Pi_s (\ell_c) \label{dWnl} \,,
\end{align}
where $\overline \Pi_s (\ell_c)$ represents {the SGS net} flux of free energy through the coarse-grained scales $\ell_c$ for the species $s$,
\begin{align}
\overline \Pi_s (\ell_c)= \int d^3R_s\int d^3v \frac{T_s}{F_s}\bigg{[}-\overline{\tau}_s \cdot\nabla\overline{h}_s \bigg{]}
\label{Flux} \,.
\end{align}%

Since the free energy is a nonlinear invariant, {\cc for a finite-scale system (as considered numerically in the current paper), we find the SGS net flux through an infinitely small coarse-grain scale to be equal to zero},
\begin{align}
\lim_{\ell_c\rightarrow 0}\frac{d\overline W}{dt}\bigg{|}_{NL}=\lim_{\ell_c\rightarrow 0} \sum_s \overline \Pi_s (\ell_c)=0 \,.
\end{align}
Furthermore, the contributions to the free energy from each plasma species are independently invariant under the action of the nonlinear terms, i.e. $\lim_{\ell_c\rightarrow 0} \overline \Pi_s (\ell_c)=0 $.
{\cc Naturally, for an infinite-scale system given by the $\mbox{Do} \rightarrow \infty$ limit, where $\mbox{Do}\sim(k_{\perp}^{\nu_i} \rho_i)^{5/3}$ is the Dorland number \citep[for the definition convention see][]{Schekochihin:2009p1131} defined here on the ion dissipation scale (i.e. the scale at which the finite collisional dissipation peaks in amplitude), taking the $\ell_c \rightarrow 0$ limit will give a constant flux value once parallel mixing can be neglected. In the current paper $Do\approx228$ and strong parallel mixing affects the scaling of the electron flux.}

With this knowledge, {we consider the {$\RR_s$}-density of free energy for each plasma species, i.e. $W_s(\RR_s,t)$, see Appendix~\ref{app_energy}, and look at its coarse-grained variation due to the action of nonlinear interactions,}
\begin{align}
\frac{\partial \overline W_s(\RR_s,t)}{\partial t}\bigg{|}_{NL}&=\int d^3v  \bigg{[}\nabla \cdot ( \overline{\bf u}_s \overline{h}_s + \overline{\tau}_s)\bigg{]}\overline{h}_s \frac{T_s }{F_s}\nonumber \\
&=\int d^3v \bigg{[} \nabla \cdot ( \overline{\bf u} \overline{h}_s^2/2+ \overline{\tau} \overline{h}_s)\frac{T_s }{F_s}\bigg{]} + \int d^3{v} \bigg{[}  -\overline{\tau}_s\cdot\nabla\overline{h}_s\frac{T_s }{F_s} \bigg{]}\, \label{Wnldens}.
\end{align}
The first term on the {\em rhs} corresponds to a transport in position space of free energy, while the second corresponds to the density of the SGS scale flux. We introduce the following definitions, 
\begin{align}
 \overline \Upsilon_s(\RR_s,t)&=\int d^3v \frac{T_s }{F_s}\bigg{[} \nabla \cdot ( \overline{\bf u}_s \overline{h}_s^2/2+ \overline{\tau}_s \overline{h}_s)\bigg{]} \label{Yidens} \ ,\\
\overline \Pi_s(\RR_s,t)&= \int d^3v \frac{T_s }{F_s} \bigg{[}  -\overline{\tau}_s \cdot\nabla\overline{h}_s \bigg{]} \label{Pidens}\ .
\end{align}
{Since the definitions above are the two integrals from eq.~(\ref{Wnldens}), we see that we could move the $\nabla \cdot ( \overline{\tau}_s \overline{h}_s)$ term from (\ref{Yidens}) to (\ref{Pidens}) in an attempt to contain the SGS contributions into a single $\overline{h}_s \nabla \cdot \overline{\tau}_s$ quantity. In fact, the SGS net flux (\ref{Flux}) is defined up to a divergence term, and the integrant could be written simply as $ \overline{h}_s \nabla \cdot \overline{\tau}_s$. However, while the SGS net flux is not changed by adding or subtracting a divergence term, the resulting spatial density would be different. It is hard to see a good reason to add an extra contribution to the flux density $\overline \Pi_s(\RR_s,t)$ that does not contribute at all to the net flux across a scale{\cc \footnote{{\cc See p.~15 of \citep{Eyink:2018p2109}, following eq.~(5.20), for a similar discussion on the entropy flux.}}}. As such, we decide on the definitions given by eqs.~(\ref{Yidens}-\ref{Pidens}) for our current work.}

{\cc
In terms of the Poisson bracket structure, the (\ref{Pidens}) integrant becomes
\begin{align}
\overline{\tau}_s \cdot \nabla \overline h_s =\frac{c}{B_0} \Big{[} \overline{ {h}_s \{\langle \chi \rangle}_{\RR_s} ,\overline{ h}_s\} - \frac{1}{2}\{ \overline{\langle \chi \rangle}_{\RR_s}, \overline{h}^2_s\} \Big{]} \label{sgsPoisson}.
\end{align}
{For clarity, 
\begin{align}
 \overline{ {h}_s \{\langle \chi \rangle}_{\RR_s} ,\overline{ h}_s\} =\overline{  {h}_s\frac{\partial \langle \chi \rangle_{\RR_s}}{\partial x}}\frac{\partial \overline{ h}_s}{\partial y}-\overline{ {h}_s\frac{\partial \langle \chi \rangle_{\RR_s}}{\partial y}}\frac{\partial \overline{ h}_s}{\partial x}
\end{align}
and it shows why the Poisson bracket notations become cumbersome when dealing with coarse graining. Only terms of the form $\overline{ \bullet \{\bullet} ,{ \bullet}\}$, that coarse grain across the Poisson bracket structure, give SGS contributions.}
From the properties of the Poisson bracket (see Appendix \ref{appA}) we know that the second term integrates spatially to zero ($x,y$ integration suffices). However, this second term is important to ensure the gauge invariance of the SGS flux density. Simple algebra shows that the transformation $h'_s=h_s+H$ and $\langle \chi' \rangle_{\RR_s}=\langle \chi \rangle_{\RR_s}+a$, with $\overline H =H$ and $\overline a=a$ leaves (\ref{sgsPoisson}) invariant. We also ask for the SGS flux density to be Galilean invariant, meaning that a change in the system of reference cannot change the intensity of turbulence, and see that the (\ref{Pidens}) definition fulfils this requirement. The link with Galilean invariance mentioned for $\overline \tau_s$ is given by ${\overline {\bf U}}={\bf U}=-\frac{c}{B_0}{[}\nabla a \times {\zz}{]}$. In the same spirit, the $h'_s=h_s+H$ invariance shows that by adding or subtracting background density values to $h_s$ (during the $\delta f$ splitting for example), we cannot change the intensity of turbulence. This also highlights why the $\overline{h}_s \nabla \cdot \overline{\tau}_s$ quantity does not make for a good SGS flux density, while eq.~(\ref{Pidens}) does.
}

We will normalise the nonlinear results in respect to,
\begin{align}
\varepsilon_{NL}=\sum_s\bigg{[}\int d^3R_s\Big{(}\Upsilon_s(\RR_s,t)+\Pi_s(\RR_s,t)\Big{)}^2\bigg{]}^{1/2} \,.
\end{align}

Since $\overline \Upsilon_s(\RR_s,t)$ is defined as the divergence of a vector field, we clearly see that it integrates to zero for periodic or appropriate asymptotic boundary conditions {(i.e. $\int  d^3R_s\, \overline \Upsilon_s(\RR_s,t) = 0 $). $\overline \Upsilon_s(\RR_s,t)$} does not contribute to the redistribution of free energy across the cut-off scale. Its role is to transport free energy in position space. The {nonlinear transport of free energy} can be seen as being due to the coarse-grained advective velocity and due to the SGS interactions, {$\overline \Upsilon_s= \overline \Upsilon_{{\bf u},s}+\overline \Upsilon_{{\tau},s}$}, with 
\begin{align}
\overline \Upsilon_{{\bf u},s}&=\int d^3v \frac{T_s }{F_s}\bigg{[} \nabla \cdot ( \overline{\bf u}_s \overline{h}_s^2/2)\bigg{]} \ ,\\
\overline \Upsilon_{{\tau},s} &=\int d^3v \frac{T_s }{F_s}\bigg{[} \nabla \cdot (\overline{\tau}_s \overline{h}_s)\bigg{]}\ .
\end{align}

\begin{figure}
\centerline{
\includegraphics[width = 0.75\textwidth]{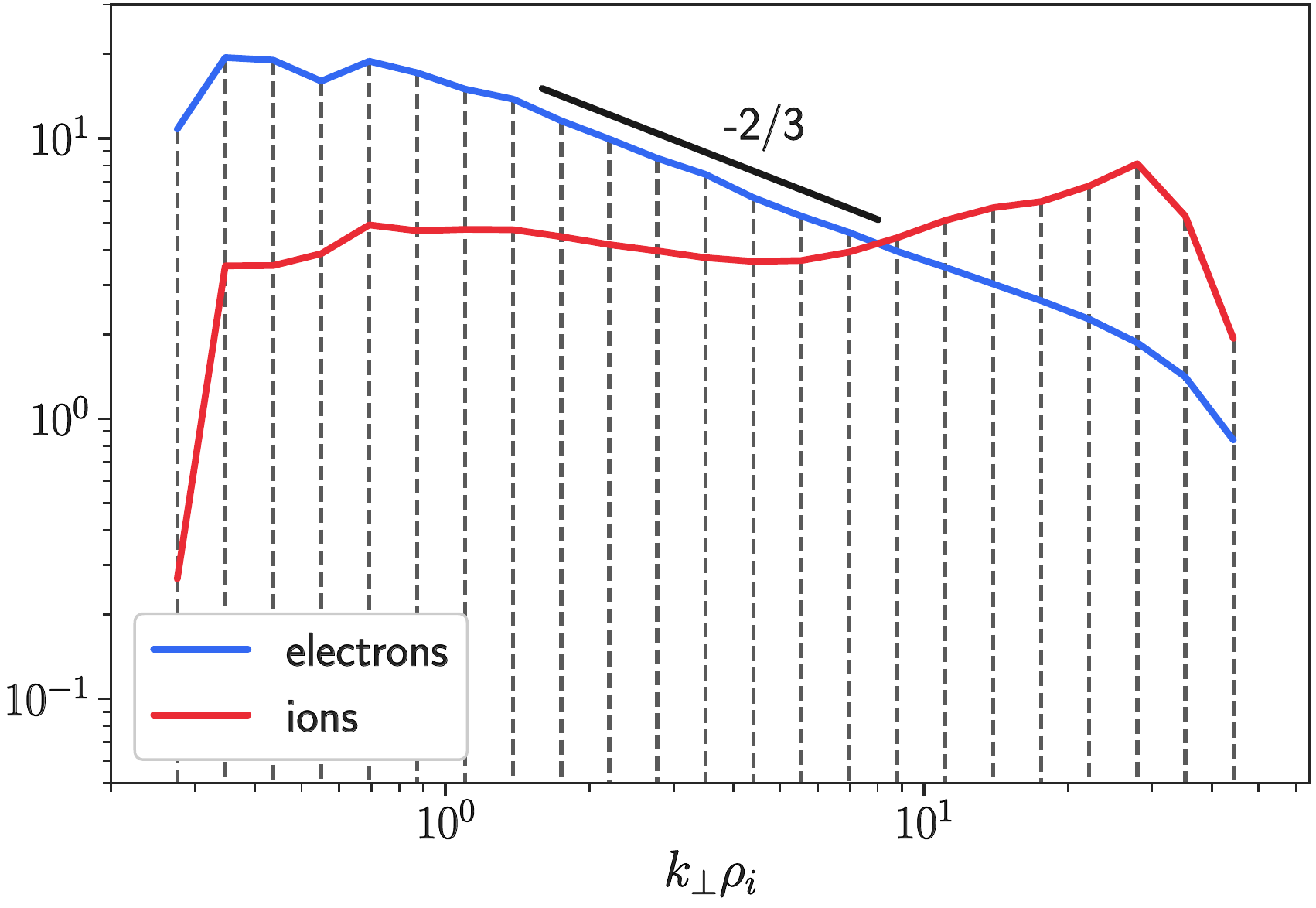}  }
\caption{{The ion and electron SGS net flux $\overline \Pi_s(\ell_c)$ normalised to $\varepsilon_{NL}$, for $\ell_c=2\pi/k_c$. The vertical dashed lines represent the same $k_c$ cut-off values} as considered in \citet{Teaca:2017p1989}. We obtain the same behaviour as previously reported in \citet{Teaca:2017p1989}, where the flux was computed via a triple-scale-decomposition. While the electron flux exhibits a $-2/3$ scaling, the ion flux tends to be constant across the "inertial range" scales. The small bottleneck around $k_\perp \rho_i\approx 26$ is due to the relatively abrupt transition towards the range of scales dominated by collisional dissipation.}
\label{fig:sgs_flux}
\end{figure} 

The density of the SGS flux is much more interesting to us. Performing the spatial integration, we recover the $\overline \Pi_s (\ell_c)$ flux,
\begin{align}
\int  d^3R_s\, \overline \Pi_s(\RR_s,t) = \overline \Pi_s(\ell_c)\,,
\end{align}
which we plot in Figure~\ref{fig:sgs_flux}. {As noted,} while the (\ref{dWnl}) integral recovers $\overline \Pi_s(\ell_c)$, it does not provide for a good definition for the SGS flux density. 

Scaling laws predicted via functional analysis, {\cc like in \cite{Eyink:2018p2109} for the Vlasov-Maxwell system}, are computed for absolute values (i.e. $L^p$ norms). This prohibits cancellation effects from occurring when integrating any sign indefinite quantity. To make a comparison, we define the maximal (upper bound) values for the spatial transfer and SGS flux. We do so by taking the absolute value before integrating the respective quantities in velocity space.  
\begin{align}
 \overline \Upsilon^{\mbox{\scriptsize \sc max}}_s(\RR_s,t)&=\int d^3v \frac{T_s }{F_s}\bigg{|} \nabla \cdot ( \overline{\bf u}_s \overline{h}_s^2/2+ \overline{\tau}_s \overline{h}_s)\bigg{|} \ ,  \label{Tmax} \\
\overline \Pi^{\mbox{\scriptsize \sc max}}_s(\RR_s,t)&= \int d^3v \frac{T_s }{F_s} \bigg{|}  \overline{\tau}_s \cdot\nabla\overline{h}_s \bigg{|}\ .\label{FluxMax}
\end{align}

Next, we present a numerical analysis of the SGS flux density and spatial transport of free energy, concentrating on one aspect at a time.

\section{Numerical analysis} \label{numanal}

\subsection{The free energy transport in position space}

\begin{figure}
\center
\includegraphics[width = 1\textwidth]{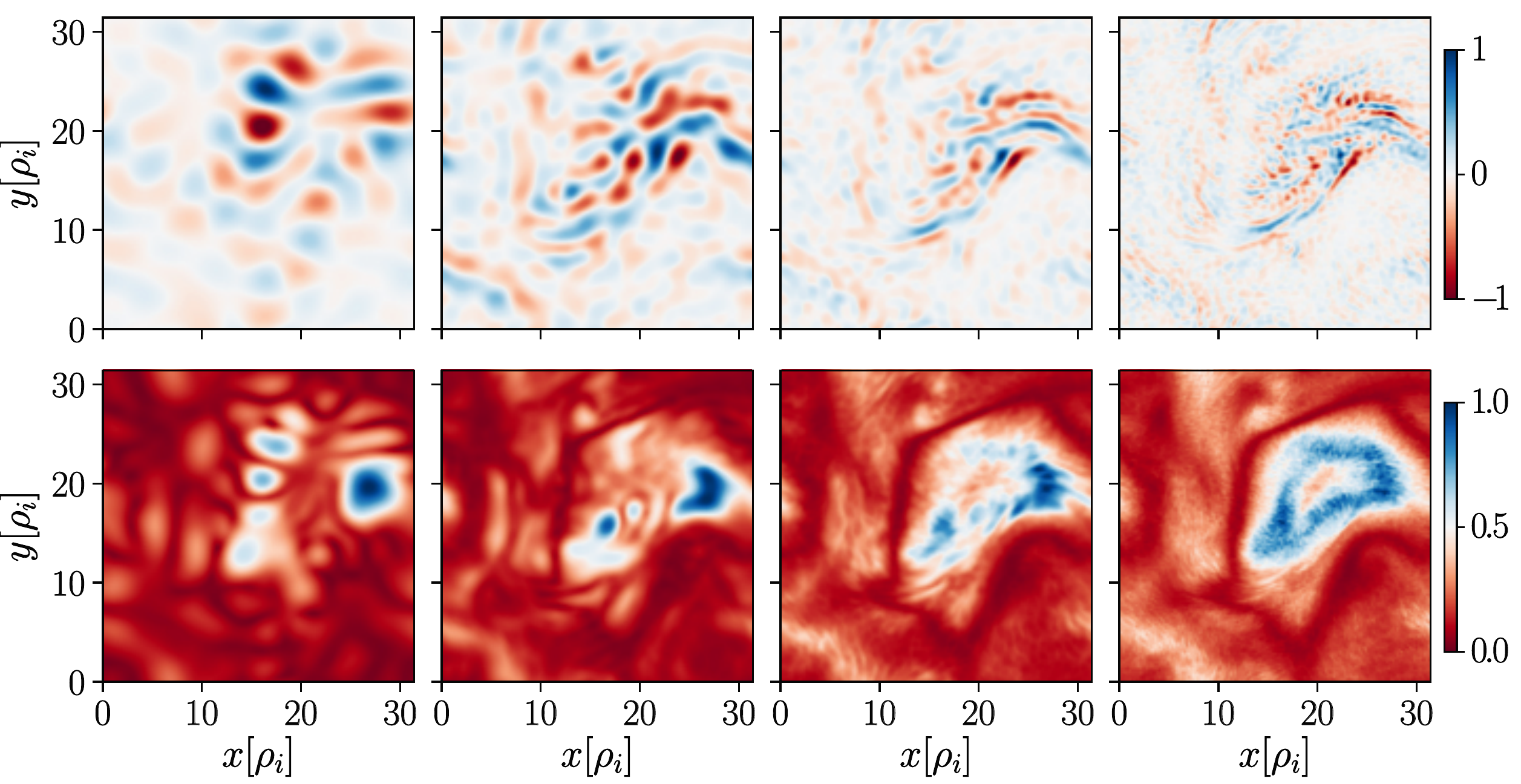}
\caption{{Top row shows} the transport of free energy in position space for the ions, {$\overline \Upsilon_i(\RR_s,t)$}. The plots {depict the typical perpendicular structures} at an arbitrary $z$ slice and are normalised to their maximal in-plane value. From left to right, the $k$-filtering cut-offs are $k_c\rho_i=\{1,2,4,8\}$. Bottom row shows {the same plots for $\overline\Upsilon^{\mbox{\scriptsize \sc max}}_i(\RR_s,t)$}.}
\label{fig:fe_trans_i}
\end{figure} 

\begin{figure}
\center
\includegraphics[width = 1\textwidth]{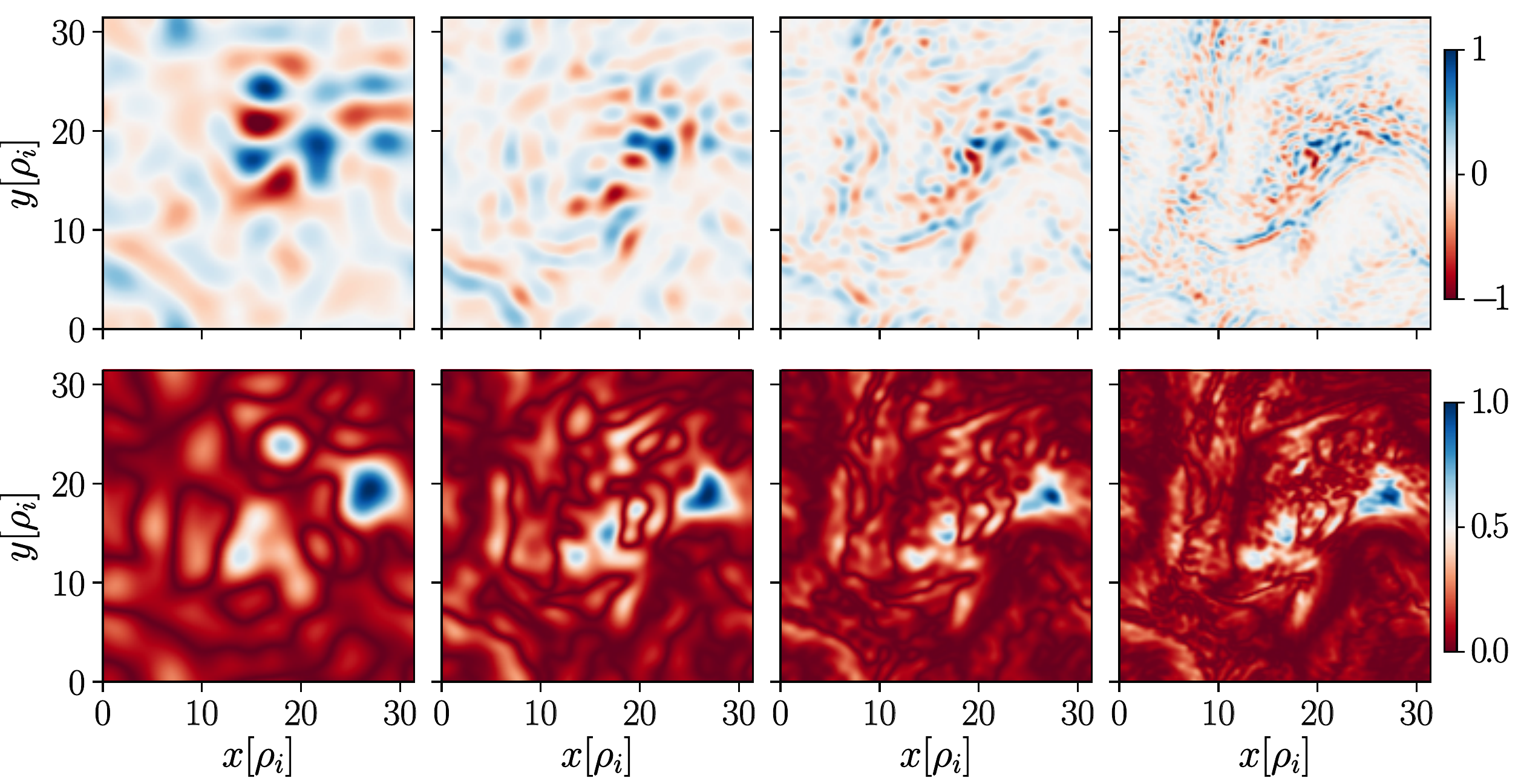}
\caption{{Top row shows} the transport of free energy in position space for the electrons, {$\overline \Upsilon_e(\RR_s,t)$}. The plots {depict the typical perpendicular structures} at an arbitrary $z$ slice and are normalised to their maximal in-plane value. From left to right, the $k$-filtering cut-offs are $k_c\rho_i=\{1,2,4,8\}$. Bottom row shows {the same plots for $\overline\Upsilon^{\mbox{\scriptsize \sc max}}_e(\RR_s,t)$}.}
\label{fig:fe_trans_e}
\end{figure} 

We plot the space density of the nonlinear transport of free energy in Figure~\ref{fig:fe_trans_i} for the ions and in Figure~\ref{fig:fe_trans_e} for the electrons, respectively. In addition, in each figure, we plot the upper bound transport density ($\overline\Upsilon^{\mbox{\scriptsize \sc max}}_s$) for the two species. Varying the cut-off value in dyadic increments (i.e. $k_c\rho_i=\{1,2,4,8\}$) allows us to observe the change in transport {as smaller and smaller structures are accounted}. For $\overline\Upsilon_s$, the cut-off scale indicates how a structure of that size perceives the spatial transport of free energy. As the cut-off scales are taken to be smaller and smaller, we see more fine-structure being added to the transport behaviour. In particular for the electrons, this is seen best from the plots of their upper bound transport ($\overline \Upsilon^{\mbox{\scriptsize \sc max}}$). For the transport, while more fine structures are added for small scales, the peaks tend not to change. {This is natural, as the advection of large scales by the small scales is negligible in most turbulent systems.}

\begin{figure}
\centerline{
\includegraphics[width = 1\textwidth]{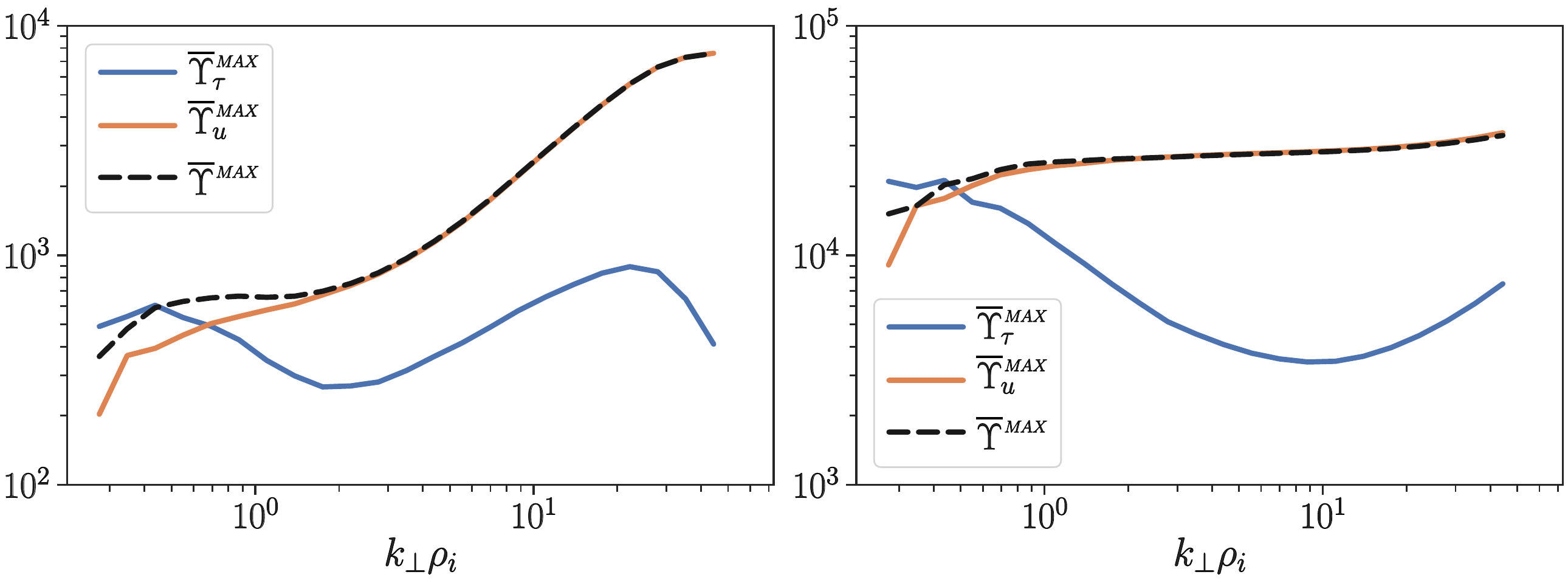} 
} 
\caption{The global variation with the {cut-off scale} of the upper bound transport of free energy ($\overline \Upsilon^{\mbox{\scriptsize \sc max}}_s$) for ions (on the left) and electrons (on the right). The individual contributions from $\overline \uu_s$ and $\overline \tau_s$ are shown as well. All the plots are normalised to $\varepsilon_{NL}$.}
\label{fig:abs_tra}
\end{figure}

\begin{figure}
\centerline{
\includegraphics[width = 0.9\textwidth]{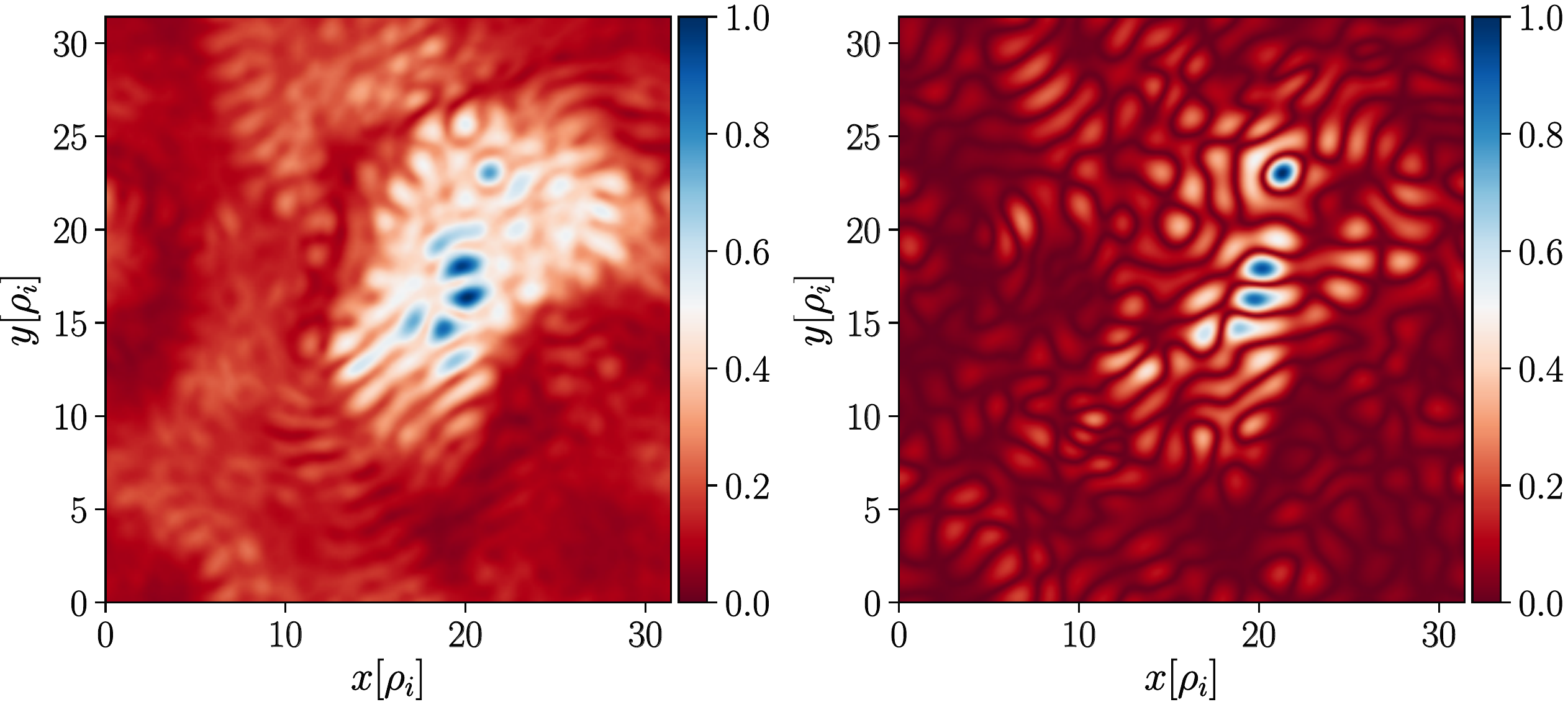}
} 
\caption{A $z$ slice in the density of $\overline \Upsilon^{\mbox{\scriptsize \sc max}}_{{\tau},i}(\RR_s,t)$ (on the left) and $\overline \Upsilon^{\mbox{\scriptsize \sc max}}_{{\tau},e}(\RR_s,t)$ (on the right) for $k_c\rho_i=2$. The plots are normalised to their maximal in-plane value. While the scale defined structures are visible as for the flux (see \S\ref{sgssec}), their overall distribution follow the $\overline \Upsilon^{\mbox{\scriptsize \sc max}}_{s}$ plots.}
\label{fig:abs_tra_dens}
\end{figure}

Since globally the spatial transport integrates to zero for any coarse-grained cut-off, we look instead in Figure~\ref{fig:abs_tra} at the global variation with scale of $\overline \Upsilon^{\mbox{\scriptsize \sc max}}_s$. Defined similarly to (\ref{Tmax}), we also plot in the same figure the upper bound values for the individual contributions {$\overline \Upsilon^{\mbox{\scriptsize \sc max}}_{{\bf u},s}=\int d^3v \frac{T_s }{F_s}\big{|} \nabla \cdot ( \overline{\bf u}_s \overline{h}_s^2/2)\big{|}$ and $\overline \Upsilon^{\mbox{\scriptsize \sc max}}_{{\tau},s}=\int d^3v \frac{T_s }{F_s}\big{|} \nabla \cdot (\overline{\tau}_s \overline{h}_s)\big{|}$}. 
As expected, doing so allows us to clearly see that the total transport is mainly due to the advective velocity and not due to the SGS terms. {\cc For ions, small-scale contributions add up fast, which we believe is due to the advective velocity and its fine perpendicular velocity structure induced by the gyroaverage, structure that cannot cancel out when taken in absolute value.} In fact, past the initial large scales, the density plots for the upper bound transport are indistinguishable from the advective velocity contribution (not shown here). For reference, we plot in Figure~\ref{fig:abs_tra_dens} a $z$-slice in the density of $\overline \Upsilon^{\mbox{\scriptsize \sc max}}_{{\tau},s}$ for the $k_c\rho_i=2$ cut-off. {Not surprising, the $\overline \Upsilon^{\mbox{\scriptsize \sc max}}_{{\tau},s}$ structures are closer in shape but not location to the ones observed for the energy flux density, as we will see next.} 

\subsection{The free energy flux density} \label{sgssec}

\begin{figure}
\center
\includegraphics[width = 1\textwidth]{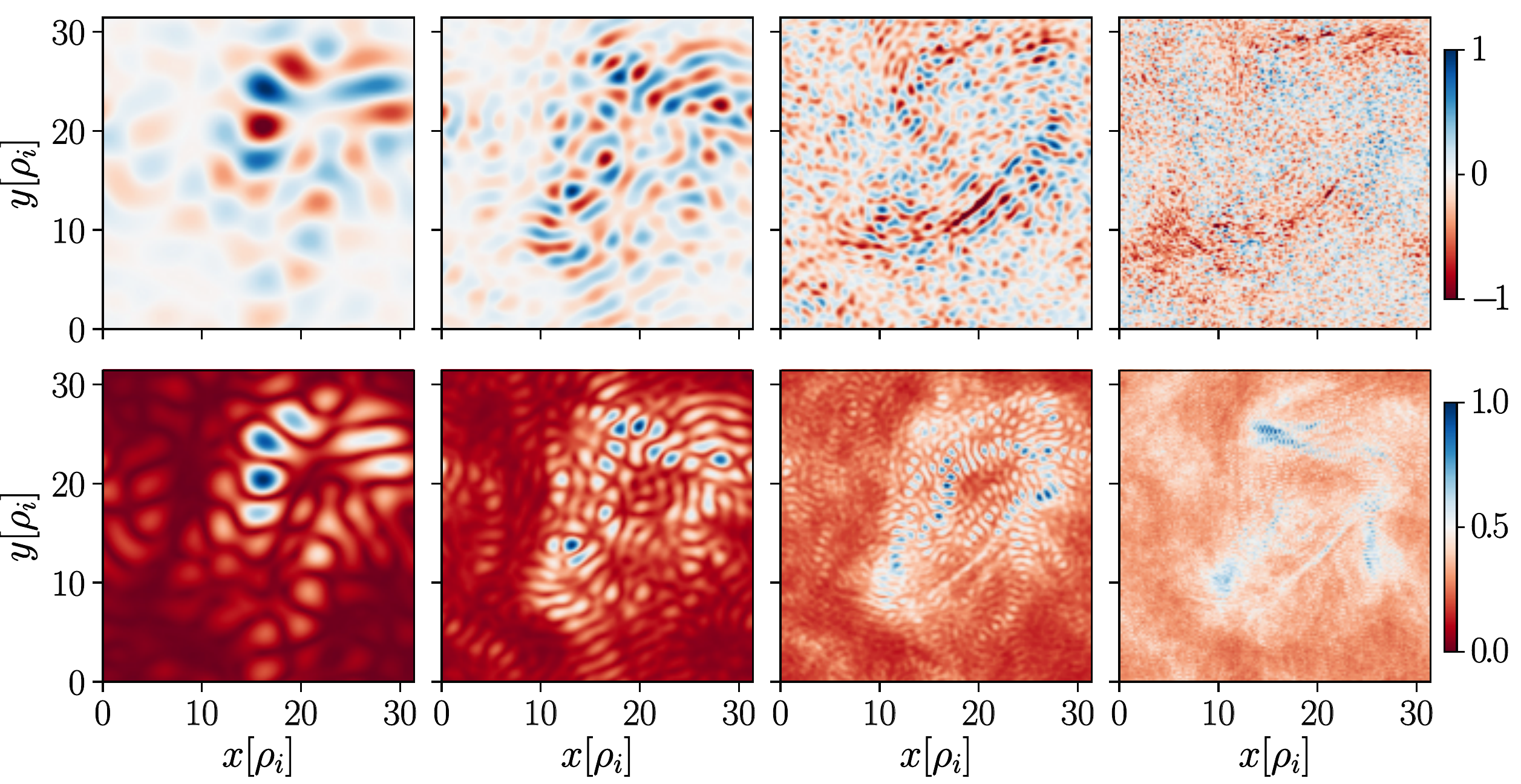}
\caption{{Top row shows} the SGS flux density of free energy for the ions, {$\overline \Pi_i(\RR_s,t)$}. The plots {depict the typical perpendicular structures} at an arbitrary $z$ point. From left to right, the $k$-filtering cut-offs are $k_c\rho_i=\{1,2,4,8\}$. Bottom row shows {the same plots for the absolute SGS flux density $\overline \Pi^{\mbox{\scriptsize \sc max}}_i(\RR_s,t)$}.}
\label{fig:fe_flux_i}
\end{figure}

\begin{figure}
\center
\includegraphics[width = 1\textwidth]{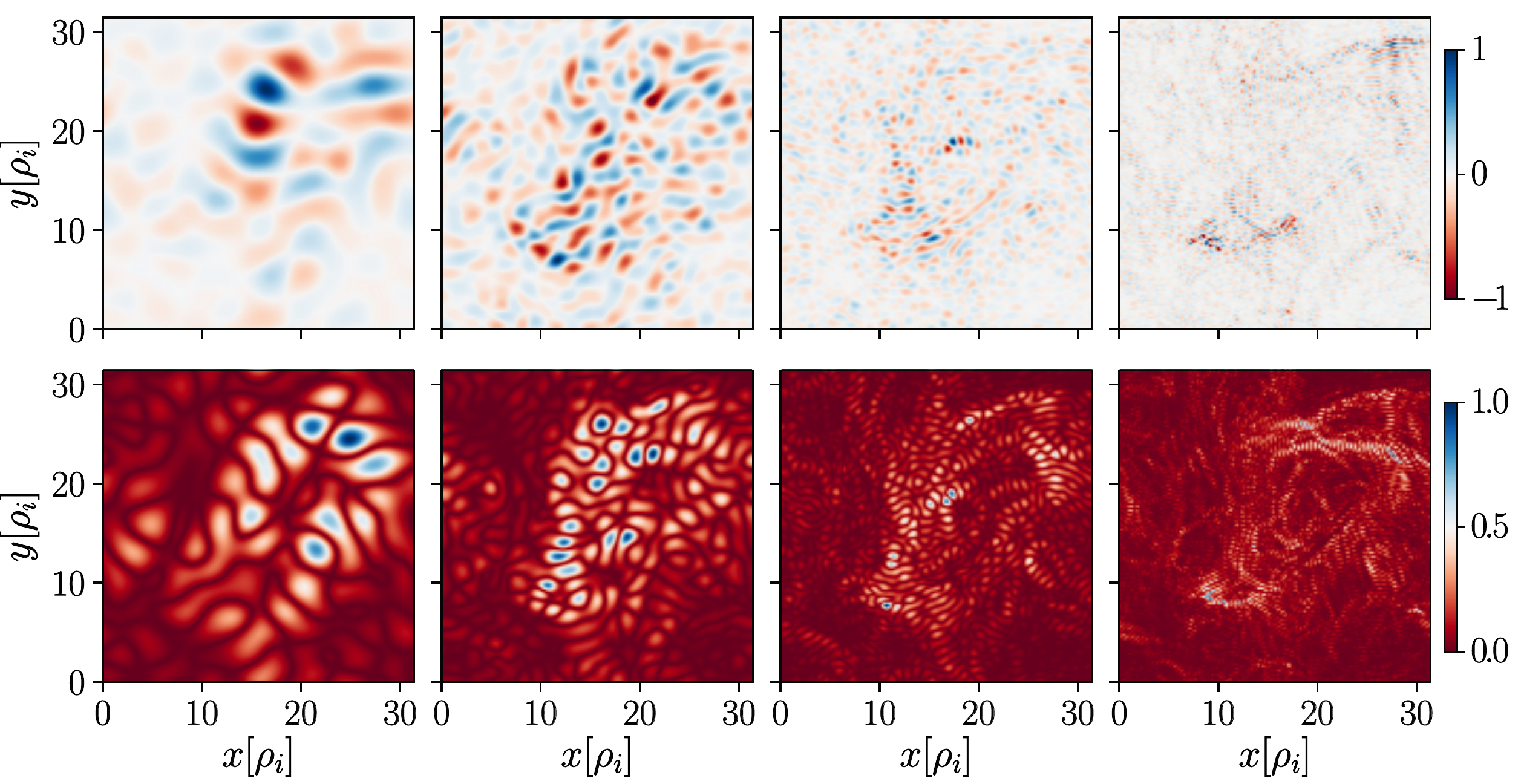}
\caption{{Top row shows} the SGS flux density of free energy for the electrons, {$\overline \Pi_e(\RR_s,t)$}. The plots {depict the typical perpendicular structures} at an arbitrary $z$ point. From left to right, the $k$-filtering cut-offs are $k_c\rho_i=\{1,2,4,8\}$. Bottom row shows {the same plots for the absolute SGS flux density $\overline \Pi^{\mbox{\scriptsize \sc max}}_e(\RR_s,t)$}.}
\label{fig:fe_flux_e}
\end{figure}

We plot the density of the SGS flux of free energy for the ions (Figure~\ref{fig:fe_flux_i}) and for the electrons (Figure~\ref{fig:fe_flux_e}), respectively. The upper bound (maximal) value of the flux density for each species are presented as well. Compared to the transport density, the SGS flux density shows that as the cut-off scales become smaller, the small scale information replaces the larger scale ones. We do not observe more fine scale structures being added on top of a larger one, but small scales replacing larger one. This is one of the best ways to perceive the flux of free energy across a scale (we will refine further this argument to account for the filter type in \S\ref{sec:kernelresults}).

\begin{figure}
\center
\includegraphics[width = 1\textwidth]{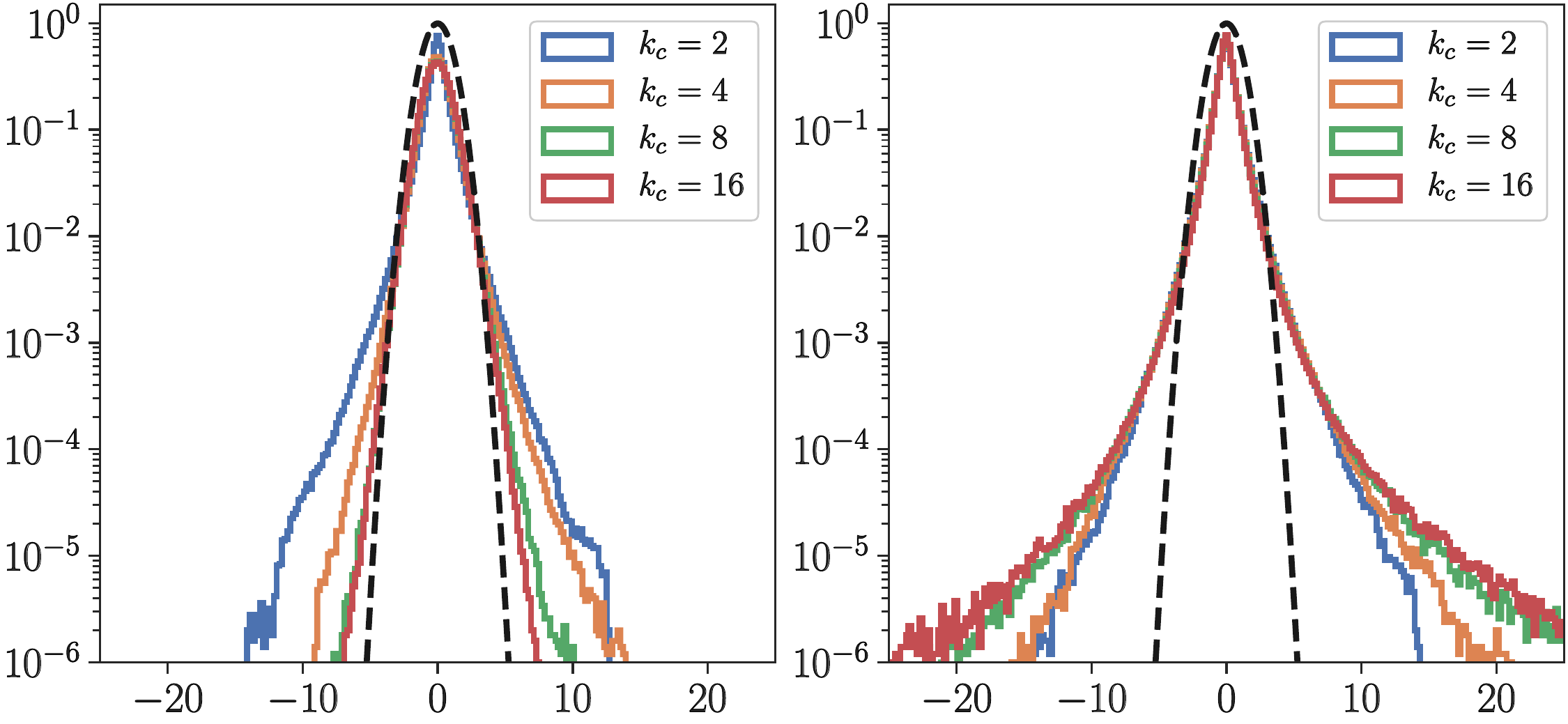} 
\caption{The normalised histogram of the SGS flux density {$\overline \Pi_s(\RR_s,t)$} for (left) ions and (right) electrons. Observe the slight asymmetry in favour of the positive values. {For visual reference, the dashed lines depict the corresponding Gaussian distributions.}}
\label{fig:fe_flux_hist}
\end{figure}

For the ions, the SGS flux density tends to homogenise for smaller and smaller structures. The electrons show an opposite behaviour, with structures of higher intensity than the background occupying a smaller and smaller volume. These behaviours are clearly seen in Figure~\ref{fig:fe_flux_hist}, where we plot the normalised histogram of the SGS flux density values. We see the histogram tails for the electrons becoming more pronounced as the cut-off scales become smaller, while the ions' values tend towards a Gaussian distribution at small scales. This is inline with the intermittency measurements performed on the distribution function in \citet{Teaca:2019p2154}.

\begin{figure}
\center
\includegraphics[width = 1\textwidth]{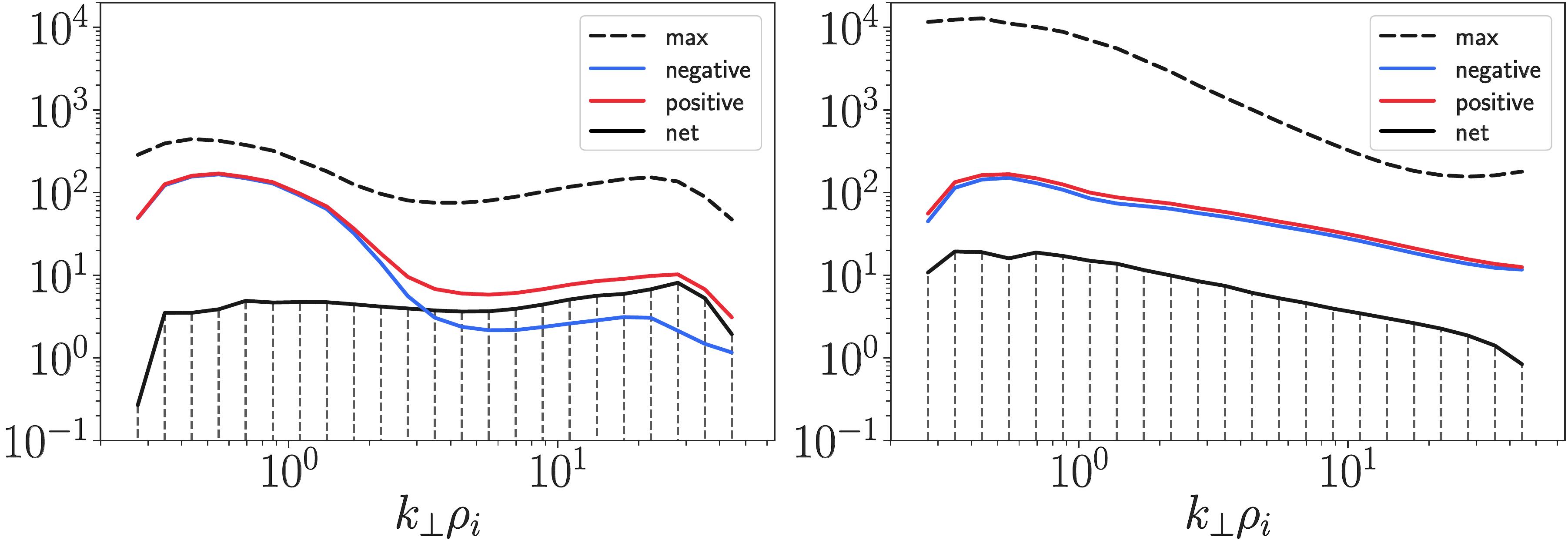} 
\caption{{The net SGS flux $\overline \Pi_s(\ell_c)$ computed as the difference of positive $\overline \Pi^{(+)}_s(\ell_c)$ (direct cascade) and negative $\overline \Pi^{(-)}_s(\ell_c)$ (backscatter) contributions for (left) ions and (right) electrons. The positive contributions dominate. For both species, the maximal flux $\overline \Pi^{\mbox{\scriptsize \sc max}}_s(\ell_c)$ given by (\ref{FluxMax}) is plotted as well. All are normalised to $\varepsilon_{NL}$. The net SGS flux and the $\ell_c=2\pi/k_c$ cut-off values depicted by vertical short-dash lines are the same as in Figure.~\ref{fig:sgs_flux}}.}
\label{fig:fe_flux_di}
\end{figure}

{One of the advantages of measuring SGS flux density is the ability to separate positive $\overline \Pi^{(+)}_s(\ell_c)$ and negative $\overline \Pi^{(-)}_s(\ell_c)$ valued contributions to the net flux,
\begin{align}
\overline \Pi^{(+)}_s (\ell_c)= &\int d^3R_s\overline \Pi_s(\RR_s,t), \ \ \mbox{for}\ \  \overline\Pi_s(\RR_s,t)>0 \ , \\
\overline \Pi^{(-)}_s (\ell_c)= -&\int d^3R_s\overline \Pi_s(\RR_s,t) , \ \ \mbox{for}\ \  \overline\Pi_s(\RR_s,t)<0 \ .
\label{FluxPM}
\end{align}}%
The positive value indicates a transfer towards the small scales, while a negative value indicates a backscatter from small scales towards large ones. From Figure~\ref{fig:fe_flux_hist}, we clearly see that the positive branch dominates. We plot in Figure~\ref{fig:fe_flux_di} the positive {($\overline \Pi^{(+)}_s$)} and negative {($\overline \Pi^{(-)}_s$)} contributions to the net flux. {The difference of the positive and negative contributions give the net flux, i.e. $\overline \Pi_s$=$\overline \Pi^{(+)}_s$-$\overline \Pi^{(-)}_s$}. {Across the entire range of scales, we} see how the net flux {for the electrons} is the result of density cancellations of an order of magnitude higher. {For the ions, a drastic cancelation is only observed up to about $k_\perp\rho_i\sim2$.} We can say that more energy is moved up and down the energy cascade in scale space than the secular-like net flux that is ultimately dissipated into heat. This is important as this diffusion in scale space has an impact on the self-organisation of turbulence. The fact that the net flux through a given scale is smaller in value than typical values of the flux density, shows the benefit of using upper bound calculations in determining the intensity of nonlinear dynamics.

\subsection{Filtering kernel dependence} \label{sec:kernelresults}

\begin{figure}
\center
\includegraphics[width = 1\textwidth]{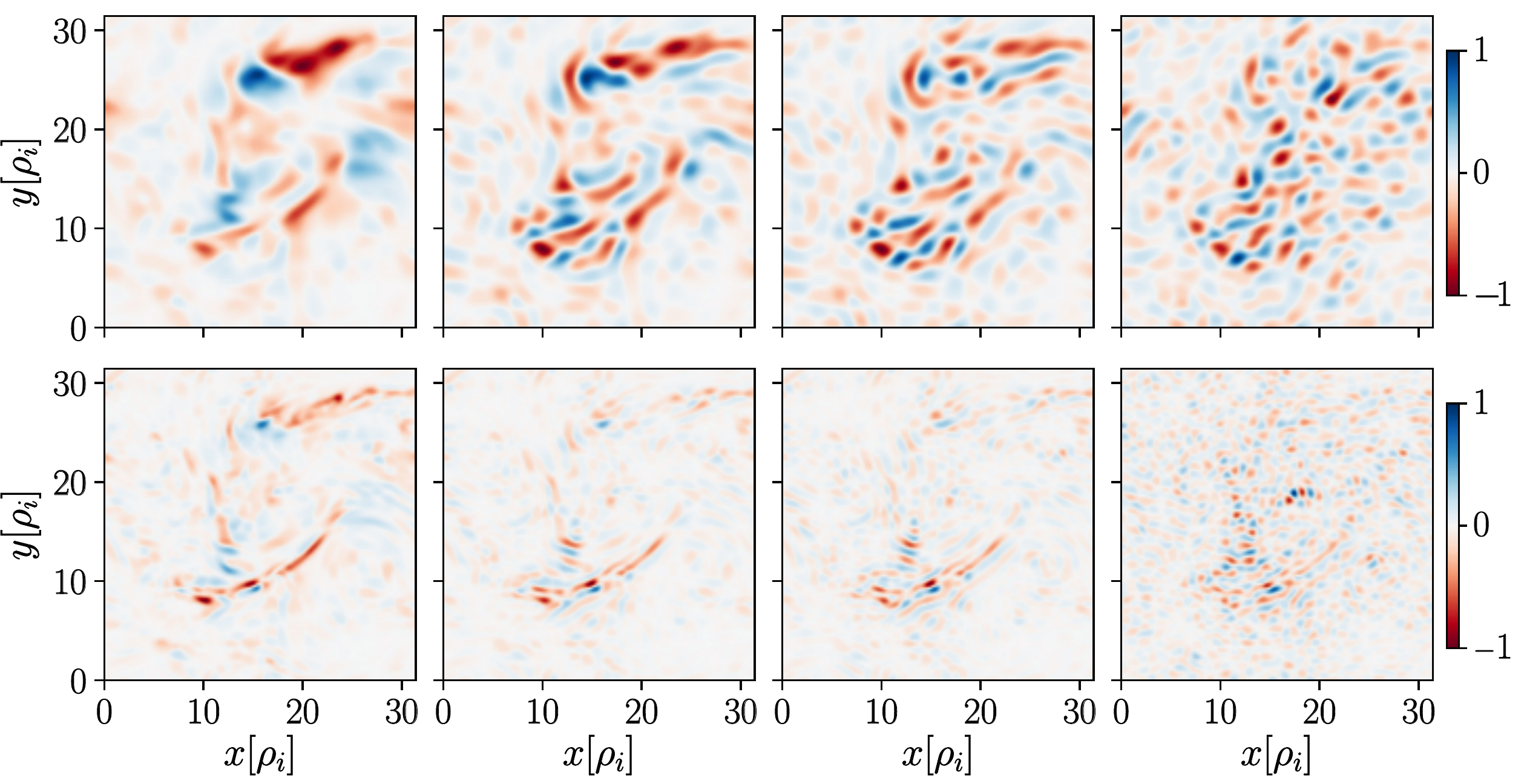}
\caption{The election's SGS flux density {$\overline \Pi_e(\RR_s,t)$} for (top) $k_c \rho_i =2$ and (bottom) $k_c \rho_i =4$. Left to right we use  $\hat G^\alpha$ with $\alpha=\{2,4,8,\infty\}$. The tendency to spatially delocalise the structures in position space, while at the same time increase the consistency of their scale size representation, can be observed for larger $\alpha$. }
\label{fig:filter_sgs_dens}
\end{figure} 

\begin{figure}
\centerline{
\includegraphics[width = 1\textwidth]{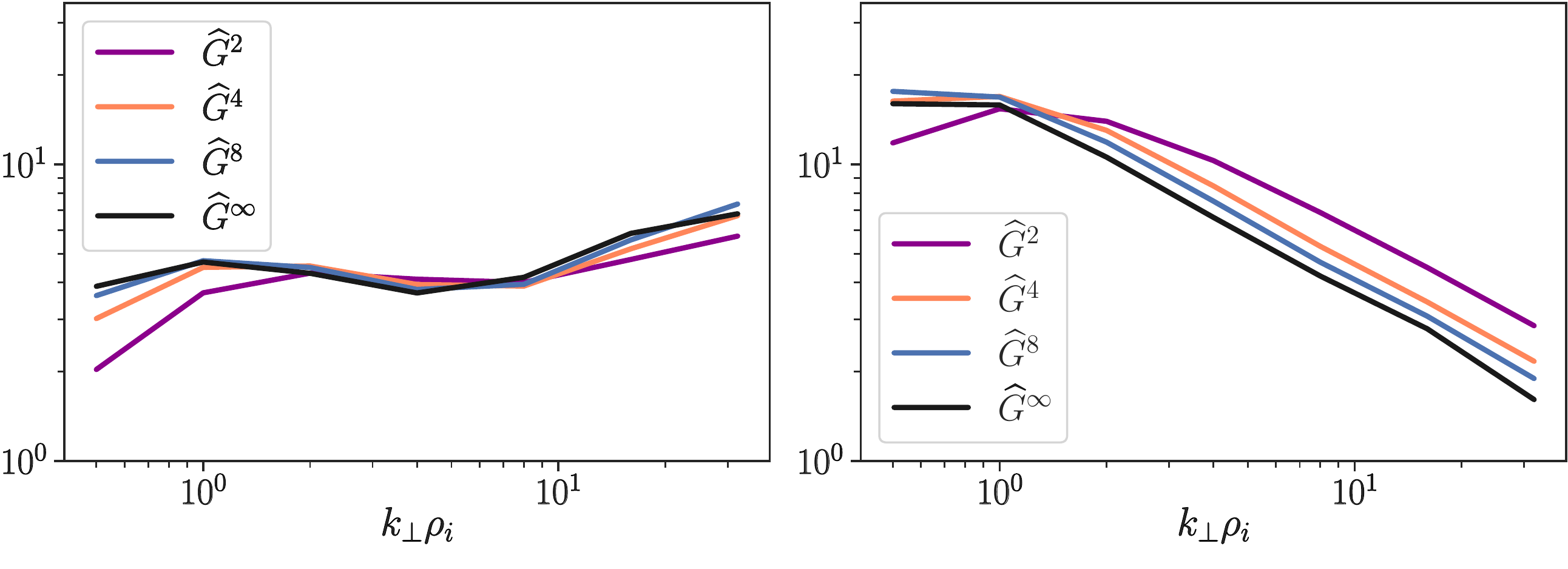} 
}
\caption{SGS net flux {$\overline \Pi_s$} for (left) ions and (right) {electrons} normalised to $\varepsilon_{NL}$, for $\hat G^\alpha$ with $\alpha=\{2,4,8,\infty\}$.}
\label{fig:filter_sgs}
\end{figure} 

\begin{figure}
\center
\includegraphics[width = 1\textwidth]{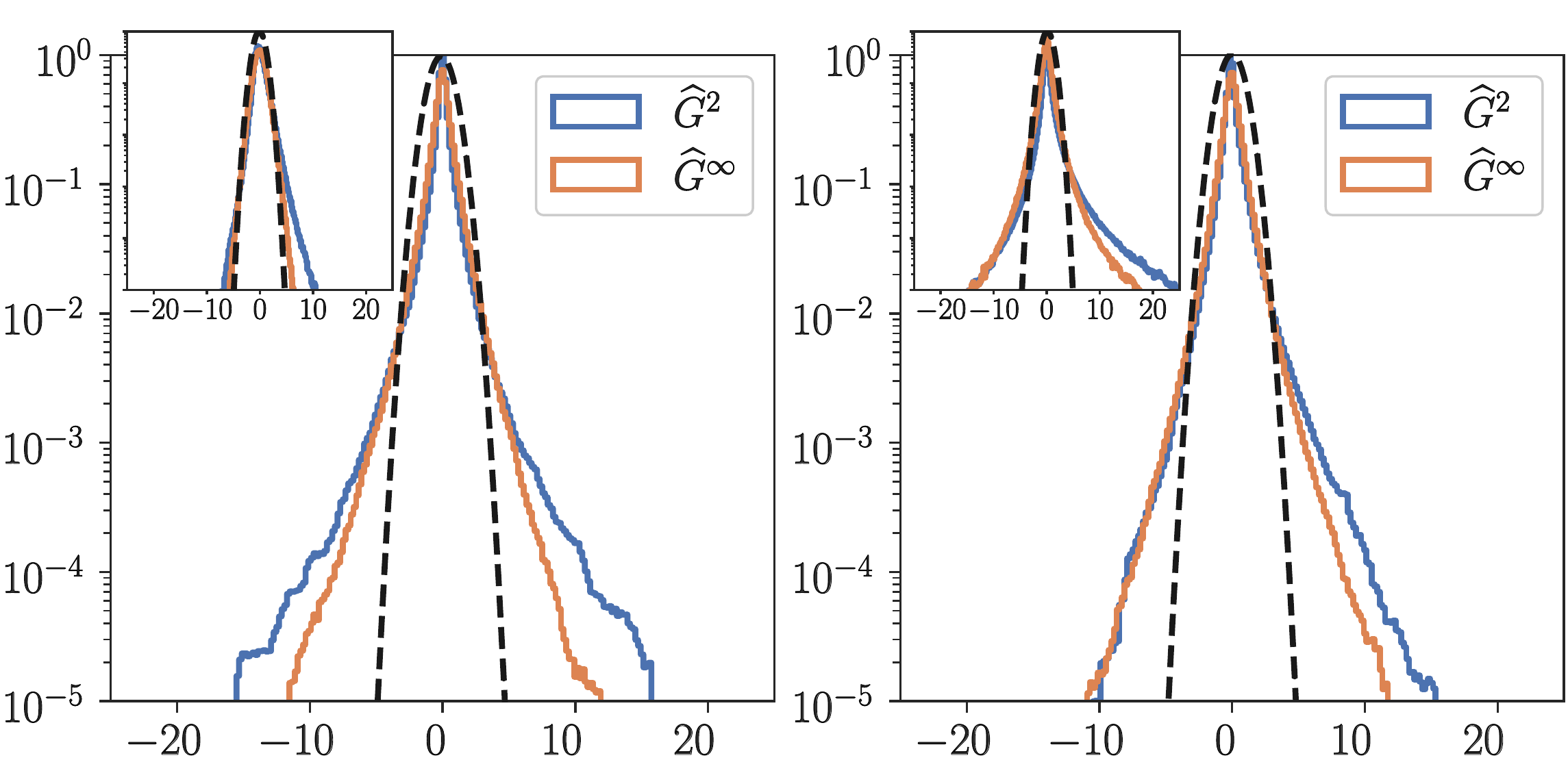} 
\caption{The normalised histogram of the SGS flux density {$\overline \Pi_s(\RR_s,t)$} at $k_c \rho_i=2$ using a Gaussian filter ($\hat G^2$) and a sharp one ($\hat G^\infty$) for (left) ions and (right) electrons. The insert pictures show the same for $k_c \rho_i=16$. We see that the Gaussian filter has increased tail contributions {and} that the positive branch (direct cascade) dominates. {For visual reference, the dashed lines depict the corresponding Gaussian distributions}. }
\label{fig:fe_flux_GScompare}
\end{figure} 

We want to understand if the filtering kernel impacts our results. In theory, the results {should be insensitive to the type of filters used}, but in practice, especially when dealing with finite resolution numerical effects, they matter. We also state that we are less concerned with the representation of the electromagnetic fields and $h_s$ as a result of the filter (we found no visual difference; not shown), and are more concerned with the change of the SGS flux and its density.

While a sharp filter can be seen as a scale separation, a Gaussian filter is best seen as separating compact structures in real space. With our choice of definition (\ref{Galpha}), we can transition from the Gaussian filter to the sharp one by increasing the value of $\alpha$. From Figure~\ref{fig:filter_sgs_dens} we see that the more compact structures of an approximate scale give way to more spread out structures of well defined scale size. The flux is shown in Figure~\ref{fig:filter_sgs}, where no change in the scaling is observed once we are past the smallest of wavenumbers. However, from Figure~\ref{fig:fe_flux_GScompare} we see that {the distribution of flux density values has more pronounced tails} for a Gaussian filter.

\section{Conclusions {and discussions}}

We revisited the problem of the redistribution of free energy in strongly magnetised plasma turbulence. The plasma is embedded in a strong straight-line magnetic guide field, and the dynamics of turbulence in the proton-electron range of scales are captured via a gyrokinetic approximation. This approximation is well suited for the analysis of the energy redistribution in phase space and the subsequent thermalisation of plasma fluctuations.  We concentrated on the redistribution of free energy in the perpendicular direction to the guide field as the result of the nonlinear interactions. Unlike past works that emphasised the spectral analysis, here, a novel approach in the field of GK turbulence was employed. For a given reference scale, we decomposed the nonlinear interactions in terms of coarse-grid-scales and sub-grid-scales. This approach {allowed} us to measure the spatial density of the SGS flux of free energy and the spatial advection of free energy. 

{Employing an appropriate definition for the SGS flux, which also accounts for its invariance to a change in the system of reference, and which recovers previously published results \citep{Teaca:2017p1989}, we were able to analyse its spatial density properties. The use of the flux density highlights the intermittent behaviour of nonlinear dynamics in turbulence, with high intensity flux structures occupying only a fraction of the total volume. For progressively smaller cut-off scales, the intermittency of the flux density increases for the electrons and decreases for the ions. This striking result, which is consistent with our previous work on phase-space intermittency \citep{Teaca:2019p2154}, should be investigated further and for a wider range of plasma parameters. }
{The dependence of filtered quantities on the type of scale filter has been analysed as well.} While a sharp filter in $k$-space provides the best scale separation, a Gaussian filter allows for better structure localisation. The hyper-Gaussian filters introduced here allowed for a transition between the sharp and Gaussian filter types. While the structures of the filtered fields do not depend strongly on the filter type, we have found that the nonlinear dynamics are sensitive enough that care needs to be shown when analysing {intermittent nonlinear} behaviour.   

We have also obtained the positive value and the negative value (backscatter) contributions to the SGS flux. The difference between the two gives the net flux across a scale, which is much smaller in value. This emphasises that nonlinear interactions are much larger in absolute amplitude than the resulting net flux, and that SGS effects should be modelled locally, at the density level. {\cc Previous studies \citep[e.g.][]{BanonNavarro:2011p1274, Nakata:2012p1387, BanonNavarro:2014p1535, Teaca:2014p1571, Maeyama:2015p1736} that studied the energetic interactions between scales, did so by looking at the coupling of spectral modes or spectral bands, meaning that they could not differentiate between contributions to a scale arising from different spatial structures. Moreover,} SGS models for LES methods that model solely the net flux significantly contaminate the local representation of structures above the cut-off scale. It is also important for SGS models to allow for the negative scale fluxes (backscatter), and move away from the idea that scale fluxes are simply sinks of energy. This is particularly important in plasma that undergoes complex self-organisation of structures at large scales, such as in tokamaks or stellarators, where global transport levels are known to be influenced by small-scale effects {\cc \citep{Gorler:2008p1617, Maeyama:2015p1736, Howard:2016p2161}. A density level approach to LES modelling would take into account that not all large  structures are equally affected.} For this, multifractal models developed for fluid flows \citep{doi:10.1063/1.1965094} can be considered. 
 {\cc Or models that account for the effect of small-scales on large-scale fluctuations can be adapted from non-equilibrium statistical physics; see \citet{Maeyama:2020p2160} on the use of the Mori-Zwanzig formalism for this purpose}. Moreover, the lessons learned from the LES modelling of {\cc passive-scalars} \citep{doi:10.1146/annurev.fluid.32.1.203} should be examined as well, since the nonlinear terms have an active and passive advection role for kinetic systems, which for GK systems is best seen from a Laguerre–Hermite representation of the equations \citep{Mandell:2018p2140}. 

Knowing that nonlinear interactions are responsible for a spatial advection of free energy in addition to the energy flux across scales, we have looked at the spatial transport of free energy. Not surprisingly, the coarse-grid-scales are found to dominate the spatial transport. This implies that while a SGS model is needed to truncate the nonlinear interaction in scale space, the coarse-grid-scale fields suffice to obtain the spatial balance of structures when investigating spatial advection. Spatial advection needs to be accounted for the analysis of saturation levels of turbulent transport, especially in complex tokamak or stellarator geometries{\cc, or in general whenever advective unstable structures develop. This also gives hope that by prescribing the large scale redistribution of free energy in position space, machine learning algorithms could be trained to identify relevant correlations between structures and guess the correct SGS density flux, providing effective SGS models in the process.}

{Last, to better understand the relation between theoretical and numerical estimates, we have computed upper-bound values for the flux and spatial transport of free energy. We have found the upper-bound (maximal) SGS fluxes to be much higher than the actual spatially integrated values that allow for cancellations. To complete our current approach for the analysis of the energy redistribution, a coarse graining of $v_\|$ scales needs to be additionally considered. This was not attempted here due to practical numerical limitations. This is a problem left for the future, that will be best performed via a drift-kinetic approximation.}   

BT would like to thank Gabriel Plunk, David Hatch and Tobias G\"orler for discussions on the theoretical and numerical of aspects of GK turbulence. This work was partially supported by B. Teaca's EPSRC grant No. EP/P02064X/1. We acknowledge the Max-Planck Princeton Center for Plasma Physics for facilitating the discussions that lead to this paper. {\cc We thank the anonymous referees for their constructive criticism, which lead to an improved form for the article.}

\appendix 

\section{Poisson bracket properties}\label{appA}

The binary operation,
\begin{align}
\big{\{}  f, g\big{\}}= [\nabla f \times \nabla g] \cdot {\zz}=\frac{\partial f}{\partial x}\frac{\partial g}{\partial y}-\frac{\partial f}{\partial y}\frac{\partial g}{\partial x}\, , \label{PB}
\end{align}
defines a Poisson bracket structure in the $x,y$ space that satisfies the properties:
\begin{itemize}

\item {\it antisymmetry}
\begin{align}
\big{\{}  f, g\big{\}}=-\big{\{}  g, f\big{\}}
\end{align}

\item {\it bilinearity}
\begin{align}
\big{\{}  \alpha f + \beta g, h\big{\}}=\alpha \big{\{} f, h\big{\}} +\beta \big{\{} g, h\big{\}} \\
\big{\{}  f, \beta g+\gamma h\big{\}}=\beta \big{\{} f, g\big{\}} + \gamma \big{\{} f, h\big{\}} 
\end{align}

\item {\it Leibniz-Newton rule}
\begin{align}
\big{\{}  fg, h\big{\}}=f\big{\{} g, h\big{\}}+\big{\{}  f, h\big{\}}g\\
\big{\{}  f, gh\big{\}}=\big{\{} f, g\big{\}}h+g\big{\{}  f, h\big{\}}
\end{align}

\item {\it Jacobi identity}
\begin{align}
\big{\{}  f,\big{\{}  g, h\big{\}}\big{\}}+\big{\{}  g,\big{\{} h, f\big{\}}\big{\}}+
\big{\{}  h,\big{\{} f, g\big{\}}\big{\}}=0
\end{align}

\item {\it null for a constant}
\begin{align}
\big{\{}  f, \alpha\big{\}}=0
\end{align}

\item {\it differential operator behaviour}
\begin{align}
D\big{\{}  f, g\big{\}}=\big{\{} Df, g\big{\}}+\big{\{}  f, Dg\big{\}}
\end{align}

\end{itemize}
The proofs for all the properties above are obtained directly from the definition (\ref{PB}) for $f,g,h$ functions of $(x,y)$ and $\alpha, \beta, \gamma$ numerical constants. In practice, the operator $D$ stands in for $\partial/ \partial t$ or $\partial/ \partial v_\|$.

From the definition, integrating by parts for appropriate boundary conditions (periodic, asymptotic, etc.) we obtain,
\begin{align}
\iint \big{\{} f, g\big{\}}dxdy=0\,.
\end{align}
As a direct consequence of bilinearity and the Leibniz-Newton rule, we obtain that the integral of the product of $\big{\{} f, g\big{\}}$ with any linear combination of $f$ and $g$ monomials is zero,
\begin{align}
\iint (\alpha f^m g^n)\big{\{} f, g\big{\}}dxdy=\frac{\alpha}{(m+1)(n+1)} \iint \big{\{} f^{m+1}, g^{n+1}\big{\}}dxdy=0\,.
\end{align}
For GK theory, this implies that since $\big{\{}  \langle \chi \rangle_{\RR_s}, h_s\big{\}}$ is the nonlinear term that leaves $h_s$ invariant (globally conserved), any statistical moments of $h_s$ (i.e. $h_s^m$) are nonlinear invariants as well. More generally, any quantity that can be written under the form of the Poisson bracket will be globally conserved.

\section{Free energy balance equation}\label{app_energy}
As presented in \citet{Howes:2006p1280, Schekochihin:2008p1034, Schekochihin:2009p1131}, starting from the GK equations,
\begin{align}
\frac{\partial h_s}{\partial t}+ \frac{c}{B_0}\big{\{} \langle \chi \rangle_{\RR_s},h_s\big{\}} + v_{\parallel}\frac{\partial h_s}{\partial z}=\frac{q_s F_{s}}{T_{s}}\frac{\partial\langle \chi \rangle_{\RR_s}}{\partial t}+\bigg{(}\frac{\partial h_s}{\partial t}\bigg{)\!}_c\, \label{B1}, 
\end{align}
multiplying by $T_s h_s/F_s$, integrating over the velocity space, position and summing over all species, we obtain 
\begin{align}
 \int d^3R_s &\sum_s \int d^3v \frac{T_s}{2F_s} \bigg{[}\frac{\partial h^2_s}{\partial t}+ \frac{c}{B_0}\big{\{} \langle \chi \rangle_{\RR_s},h^2_s\big{\}} + v_{\parallel}\frac{\partial h^2_s}{\partial z}\bigg{]}\nonumber\\
&= \int d^3R_s \sum_s \int d^3v {q_s h_s}\frac{\partial\langle \chi \rangle_{\RR_s}}{\partial t}+\int d^3R_s\sum_s \int d^3v\frac{T_s}{F_s}h_s\bigg{(}\frac{\partial h_s}{\partial t}\bigg{)\!}_c\ \label{B2}\ .
\end{align}
Defining $d\ast/dt= {\partial \ast}/{\partial t}+ \frac{c}{B_0}\big{\{} \langle \chi \rangle_{\RR_s}, \ast\big{\}} + v_{\parallel}{\partial \ast}/{\partial z}$ in the gyrocenter space, we write the {\it lhs} term as,
\begin{align}
\int d^3R_s &\sum_s\int d^3v \frac{T_s}{2F_s}\bigg{(} \frac{\partial h^2_s}{\partial t}+ \frac{c}{B_0}\big{\{} \langle \chi \rangle_{\RR_s},h^2_s\big{\}} + v_{\parallel}\frac{\partial h^2_s}{\partial z} \bigg{)}\nonumber \\
 &=\frac{d}{dt}\int d^3R_s \sum_s \int d^3v \frac{T_s  h_s^2}{2F_s} =\frac{d}{dt}\int d^3r \sum_s \int d^3v \frac{T_s \langle h_s^2 \rangle_{\rr}}{2F_s} \,.
\end{align}
On the {\it rhs}, using $\chi= \phi  -\vv \cdot {\bf A}/{c}$, we manipulate the first term as, 
\begin{align}
\int d^3R_s \sum_s &\int d^3v\,q_s \frac{\partial\langle \chi \rangle_{\RR_s}}{\partial t}h_s=\int d^3R_s \sum_s q_s  \int d^3v\,  \bigg{ \langle} \frac{\partial \chi }{\partial t} h_s \bigg{\rangle}_{\RR_s} \nonumber \\
&=\int d^3r \sum_s q_s  \int d^3v \bigg{ \langle} \frac{\partial \chi }{\partial t} h_s \bigg{\rangle}_{\rr}\nonumber \\
&=\int d^3r \sum_s q_s  \int d^3v \bigg{ \langle} \frac{\partial \phi }{\partial t} h_s - \frac{1}{c}\frac{\partial {\bf A} }{\partial t}\cdot  \vv\, h_s \bigg{\rangle}_{\rr}\nonumber \\
&=\int d^3r \sum_s q_s  \int d^3v \bigg{ \langle} \frac{d \phi }{d t} h_s - \nabla_\rr \phi \cdot \vv\, h_s- \frac{1}{c}\frac{\partial {\bf A} }{\partial t}\cdot  \vv\, h_s \bigg{\rangle}_{\rr}\nonumber \\
&=\int d^3r \sum_s q_s  \int d^3v \bigg{ \langle} \frac{d \phi }{d t} h_s + {\bf E}\cdot  \vv\, h_s \bigg{\rangle}_{\rr}\nonumber \\
&=\int d^3r  \frac{d \phi }{d t} \sum_s q_s  \int d^3v \big{ \langle} h_s \big{\rangle}_{\rr}  +  \int d^3r {\bf E}\cdot  \sum_s q_s  \int d^3v \big{ \langle}  \vv\, h_s \big{\rangle}_{\rr}\nonumber \\
&=  \int d^3r  \frac{d \phi }{d t} \sum_s \phi  \frac{q^2_s n_s}{2T_s} +  \int d^3r {\bf E}\cdot {\bf j}\nonumber \\
&= \frac{d }{d t} \bigg{[}\int d^3r   \sum_s \frac{q^2_s \phi^2 n_s}{2T_s} - \int d^3r \frac{B^2}{8\pi}\bigg{]}\,,
\end{align}
where we have used the relation $d \phi /dt=\partial  \phi /\partial t+\vv \cdot \nabla_\rr \phi$, the electric field definition ${\bf E}=-\nabla_\rr \phi + \partial A/c\partial t$ and the electric current expression ${\bf j}=\sum_s q_s  \int d^3v \big{ \langle}  \vv\, h_s \big{\rangle}_{\rr}$ . For the last equality we have used {the quasi-neutrality condition} $\sum_s q_s \int \langle h_s\rangle_{\rr} d^3v=\sum_s q_s\frac{q_s \phi}{T_s}n_s$ and the Poynting theorem in the form, 
\begin{align}
 \int d^3r {\bf E}\cdot {\bf j}=-  \frac{d }{d t} \int d^3r \frac{B^2}{8\pi}\, .
\end{align}

Grouping all the terms and knowing that the last term on the {\it rhs} represents the change of free energy due to collisions, we obtain the free energy balance equation,
\begin{align}
\frac{dW}{dt}&=\frac{d}{dt}\int d^3r\,\bigg{[} \sum_s\bigg{(}\int d^3v \frac{T_s \langle h_s^2 \rangle_{\rr}}{2F_s} - \frac{q^2_s\phi^2n_s}{2T_s}\bigg{)}+\frac{B^2}{8\pi}\bigg{]}\nonumber\\
&=\int d^3R_s\sum_s \int d^3v\frac{T_s}{F_s}h_s\bigg{(}\frac{\partial h_s}{\partial t}\bigg{)\!}_c\ .
\end{align}

{Last, we define the $\RR_s$-density of the free energy contribution of species $s$ as  
\begin{align}
W_s(\RR_s,t)=  \int d^3v \bigg{[}h_s- \frac{q_s F_{s}}{T_{s}}\langle \chi \rangle_{\RR_s}\,\bigg{]}\frac{T_s}{F_s}h_s,
\end{align}
The quantity $W_s(\RR_s,t)$ recovers the free energy upon summing over the plasma species and integrating over the position space. To show this, one just needs to trivially follow the steps presented in this appendix. From (\ref{B1}), multiplying by $T_s h_s/F_s$ and integrating only over the velocity space, we find the balance equation for $W_s(\RR_s,t)$ to be
\begin{align}
\frac{\partial W_s(\RR_s,t)}{\partial t} = \int d^3v \frac{T_s}{2F_s} \bigg{[}- \frac{c}{B_0}\big{\{} \langle \chi \rangle_{\RR_s},h^2_s\big{\}} - v_{\parallel}\frac{\partial h^2_s}{\partial z} + 2h_s\bigg{(}\frac{\partial h_s}{\partial t}\bigg{)\!}_c\ \bigg{]}.
\end{align}
We clearly see now that the variation of the free energy density for each species is due to the actions of a nonlinear term, a linear parallel term and a collisional term.}


\end{document}